\documentclass[preprint,11pt,numbers,sort&compress]{elsarticle}
\usepackage{geometry}
\usepackage{comment}
\usepackage{amsmath}
\usepackage{amsfonts}
\usepackage{amssymb}
\usepackage{hyperref}
\usepackage{mmacells}
\usepackage[varqu,varl,var0,scaled=0.97]{inconsolata}
\usepackage{soul}
\NewDocumentCommand \qpw{m} {q_{#1,0}^{(+)}}

\def\Mathematica{{{\sc Mathematica}}}
\def\Fiesta{{{\sc Fiesta 4.2}}}
\def\SecDec{{{\sc SecDec 3.0}}}
\def\FiniteFlow{{{\sc FiniteFlow}}}
\def\Lotty{{{\sc Lotty}}}

\usepackage{tikz}
\usetikzlibrary{"arrows", "automata", "backgrounds", "calendar", "chains", "matrix", "mindmap", "patterns", "petri", "shadows", "shapes.geometric", "shapes.misc", "spy", "trees"}
\usetikzlibrary{arrows,shapes}
\usetikzlibrary{trees}
\usetikzlibrary{matrix,arrows} 	
\usetikzlibrary{positioning}				% For "above of=" commands
\usetikzlibrary{calc,through}				% For coordinates
\usetikzlibrary{decorations.pathreplacing}
\usepackage{pgffor}

\usetikzlibrary{decorations.pathmorphing}	% For Feynman Diagrams
\usetikzlibrary{decorations.markings}
\tikzset{
    fermion/.style={draw=black, postaction={decorate},
        decoration={markings,mark=at position .52 with {\arrow[draw=black]{>}}}},
    fermionbar/.style={draw=black, postaction={decorate},
        decoration={markings,mark=at position .5 with {\arrow[draw=black]{<}}}},
    fermionnoarrow/.style={draw=black},
    phi/.style={dashed, draw=red}
}

\newcommand{\semiloop}[4][]{%
        \draw[#1] let \p1 = ($(#3)-(#2)$) in (#3) arc (#4:({#4+180}):({0.5*veclen(\x1,\y1)});)
}

\hypersetup{
colorlinks=true,
linkcolor=blue,
linktocpage=true,
citecolor=violet
}
\usepackage[absolute]{textpos}
\allowdisplaybreaks

\newcommand{\munich}{Max-Planck-Institut f\"ur Physik, Werner-Heisenberg-Institut, 
80805 M\"unchen, Germany.}

\begin{document}
\journal{the European Physical Journal C}
%\begin{comment}
\begin{textblock*}{100ex}(1\textwidth,5ex)
{MPP-2021-11}
\end{textblock*}

\title{\Lotty{} -- The loop-tree duality automation}
\author[a]{William J. Torres~Bobadilla\texorpdfstring{\corref{author1}}{}}
\cortext[author1]{\href{mailto:torres@mpp.mpg.de}{torres@mpp.mpg.de}}
\address[a]{\munich}

\thispagestyle{empty}
\begin{abstract}
Elaborating on the novel formulation of the loop-tree duality,
we introduce the \Mathematica{} package \Lotty{} that automates the latter
at multi-loop level. By studying the features of \Lotty{}
and recalling former studies, we discuss
that the representation of any multi-loop amplitude 
can be brought in a form, at integrand level, that only displays physical information,
which we refer to as the causal representation of multi-loop Feynman integrands.
In order to elucidate the role of \Lotty{} in this automation, 
we recall results obtained for the calculation of the dual representation 
of integrands up-to four loops. 
Likewise, within \Lotty{} framework, 
we provide support to the all-loop causal representation recently
conjectured by the same author. 
The numerical stability of the integrands generated by \Lotty{} is 
studied in  two-loop planar and non-planar topologies, 
where a numerical integration is performed and compared with 
known results. 

\end{abstract}

\maketitle
\tableofcontents

\newpage

\setlength{\parskip}{0.5ex}
\section{Introduction}

The predictions of physical observables at high energies through the perturbative 
approach of Quantum Field Theories has drawn the attention of the 
scientific community when trying to understand in more details the physics 
at experiments. In fact, a support of the theory on the experimental side
needs to be provided in view of the synergy  with
the collider machines, such as LHC at CERN~\cite{Mangano:2020icy},
and the future colliders that aim higher luminosity~\cite{Abada:2019lih,Abada:2019zxq,Abada:2019ono,Benedikt:2018csr,Bambade:2019fyw,Djouadi:2007ik,Roloff:2018dqu,CEPCStudyGroup:2018ghi,Blondel:2019vdq,Banerjee:2020tdt}. 
Thus, to calculate these observables, we rely on scattering 
amplitudes, which turn out to be the backbone in these predictions. 

Several approaches to compute scattering amplitudes at tree and multi-loop level
have been developed from the physical and mathematical properties of the latter~\cite{Bern:1994zx,Bern:1995db,Britto:2004nc,Ellis:2007br,Ossola:2006us,Mastrolia:2011pr,Badger:2012dp,Zhang:2012ce,Mastrolia:2012an,Mastrolia:2012wf,Ita:2015tya,Mastrolia:2016dhn,Mastrolia:2016czu}.
In fact, leading (LO) and next-to-leading (NLO) order predictions have been automated 
in different frameworks~\cite{Hahn:1998yk,vanHameren:2009dr,Bevilacqua:2011xh,Berger:2008sj,Hirschi:2011pa,Cascioli:2011va,Badger:2012pg,Peraro:2014cba,Cullen:2014yla,vanHameren:2010cp,Actis:2016mpe,Denner:2016kdg,Patel:2016fam,Carrazza:2016gav}, allowing to compare in many details 
the data delivered by experiments against theoretical predictions. 
Since a perturbative approach is followed, it is clear that aiming at 
high accuracy implies increasing the complexity of calculations. 
In particular, the computation of scattering amplitudes through Feynman diagrams
becomes cumbersome when the multiplicity of external legs as well as
the loop order start increasing. 

Despite the plethora of diagrams that need to be considered when going 
to higher orders, new approaches,  based on clever mathematical ideas, 
are currently being proposed to increase the efficiency and reduce the computing time 
in the several calculations~\cite{Chetyrkin:1981qh,Laporta:2001dd,Larsen:2015ped,Bendle:2019csk,Mastrolia:2018uzb,Frellesvig:2019kgj,Frellesvig:2019uqt,Weinzierl:2020xyy}.
%On top of these ideas, the automation of these techniques have allowed, to start, 
%the so-called NLO revolution, 
%where many results that were thought to be unattainable were achieved.
Then, with these ideas and methods at work, the clear motivation is extending 
the knowledge of NLO to higher orders, N...NLO.
However, the transition has not been straightforward and many subtleties
that were not present at NLO have appeared.
For instance, the evaluation of multi-loop Feynman integrals is in fact
an open problem, since, differently to the one-loop case, 
a complete basis of master integrals is not known and yet
need to be computed~\cite{Kotikov:1991pm,Henn:2013pwa,Argeri:2014qva,Borowka:2015mxa,Smirnov:2015mct}.
Nonetheless, calculations, at two-loop level,  
including five external partons, have started to appear
displaying relevant results for the physics at LHC~\cite{Badger:2019djh,Chawdhry:2019bji,Caola:2020dfu,Kallweit:2020gcp,Badger:2021nhg,Agarwal:2021grm,Badger:2021owl}.

Hence, to provide an alternative approach to deal with the 
numerical evaluation of multi-loop scattering amplitudes, 
we elaborate on the loop-tree duality (LTD) formalism, 
originally developed in Refs.~\cite{Catani:2008xa,Bierenbaum:2010cy,Bierenbaum:2012th,Buchta:2014dfa,Buchta:2015wna},
and successfully employed in applications at one~\cite{Jurado:2017xut,Driencourt-Mangin:2017gop,Aguilera-Verdugo:2019kbz,Plenter:2020lop} and two~\cite{,Driencourt-Mangin:2019aix,Driencourt-Mangin:2019yhu} loops. 
The main idea of LTD, together with its motivation, is to open loops into connected
trees, in such way that all contributions, independently of the loop order, 
are equally treated.
This is, for instance, the case of the four-dimensional unsubtracted (FDU) scheme~\cite{Hernandez-Pinto:2015ysa,Sborlini:2016gbr,Sborlini:2016hat},
which treats virtual one-loop with real tree-level diagrams as phase space integrals.
This is carried out, in order to locally cancel, before integrating, infrared (IR) and ultraviolet (UV)
singularities. 
Apart from the interplay of LTD with FDU, other formalisms that aim at having local 
cancellations of IR and UV singularities as well as a purely four-dimensional representation 
of the space-time dimension are currently being extended to have a complete automation at 
N$^2$LO~\cite{Pittau:2012zd,Donati:2013iya,Fazio:2014xea,Primo:2016omk,Mastrolia:2015maa,Hernandez-Pinto:2015ysa,Sborlini:2016gbr,Sborlini:2016hat,Capatti:2020xjc,Prisco:2020kyb}. 
For recent reviews, we refer the reader to Refs.~\cite{Gnendiger:2017pys,Heinrich:2020ybq,TorresBobadilla:2020ekr}. 

In the spirit of harnessing from the knowledge of LTD, a novel formulation of the latter 
was recently provided in Ref.~\cite{Verdugo:2020kzh}, where the approach to obtain a dual representation
of multi-loop Feynman integrals,
within the LTD framework, allowed for a complete automation and analysis regardless 
of the loop order, e.g. the works of Refs.~\cite{Aguilera-Verdugo:2020kzc,Ramirez-Uribe:2020hes,Aguilera-Verdugo:2020nrp,Bobadilla:2021rmu}
perform an analysis of up-to four loop topologies. 
Besides the generation of dual integrands, 
we have recently observed that LTD satisfies a representation free of 
unphysical singularities when all dual integrands are summed up. 
For the latter we refer to the causal representation of Feynman integrands,
since these integrands are expressed in terms of causal (physical) thresholds. 
Explicit results, together with numerical evaluations, 
for topologies up-to three~\cite{Aguilera-Verdugo:2020kzc}
and four~\cite{Ramirez-Uribe:2020hes} loops were provided 
by reconstructing their analytic expression over finite fields~\cite{vonManteuffel:2014ixa,Peraro:2016wsq,Peraro:2019svx,Klappert:2019emp}. 
To this end, we profited on the available tools and made use 
of the \FiniteFlow{} framework~\cite{Peraro:2019svx}. 
We remark that, apart from our novel representation, 
alternative approaches to the LTD formalism have been addressed
in Refs.~\cite{Tomboulis:2017rvd,Runkel:2019zbm,Runkel:2019yrs,Capatti:2019edf,Capatti:2019ypt,Capatti:2020ytd,Sborlini:2021owe}. 
In particular, let us remark the implementation of Ref.~\cite{Capatti:2020ytd} 
with its stand-alone code \texttt{cLTD.py}.

In view of the above-mentioned features of the novel formulation of LTD, 
it is desirable to provide a pedestrian code to study in more details 
how the dual integrands and, therefore, the causal representation is generated within this framework. 
In fact, the algorithm proposed in Ref.~\cite{Verdugo:2020kzh}, considered firstly 
as the iterative application of the Cauchy residue theorem has been then extended, and mathematically supported 
in Ref.~\cite{Aguilera-Verdugo:2020nrp}, where the treatment of dual integrands is modified by the 
nested application of the residue. 
Having in mind these features, we constructed a \Mathematica{} package that automates the 
extraction of these residues by following the lines of Refs.~\cite{Verdugo:2020kzh,Aguilera-Verdugo:2020nrp},
the LOop-Tree dualiTY automation (\Lotty), and its presentation is the main target 
of this paper. 
In effect, several conjectures of results obtained for up-to four loop topologies
have always been cross-checked under the umbrella of \Lotty. 

Furthermore, in a recent work~\cite{Bobadilla:2021rmu}, we generalise the findings of Ref.~\cite{Aguilera-Verdugo:2020kzc,Ramirez-Uribe:2020hes}
and conjectured a close formula for the all-loop causal representation without performing
neither an actual application of LTD nor a reconstruction over finite fields. 
In order to give a support to this conjecture, we implement in \Lotty{}  the close 
formula and give the procedure to generate the causal representation 
of any loop topology from its features. 
In fact, we remark that any topology is 
characterised by the number of cusps and the connections among them, edges.
Specifically, the knowledge of causal thresholds can be understood by 
only considering the structure of the topology, which clearly depends 
on the number of cusps and edges. 

\smallskip
This paper is organised as follows. 
In Sec.~\ref{sec:ltd1}, we briefly discuss the LTD formalism and set the notation
used throughout this work. Then, in Sec.~\ref{sec:ltd0}, we introduce 
\Lotty{} and explain how to extract dual integrands,
where, to elucidate the discussion, 
we recall the results for three and four loops found in Refs.~\cite{Verdugo:2020kzh,Ramirez-Uribe:2020hes}.
We elaborate on the causal representation, in Sec.~\ref{sec:causal}, 
by presenting the built-in routines in \Lotty{} to symbolically extract causal thresholds
and also generate the causal representation of a loop topology from its
cusps and edges. 
As an application, we explicitly show the causal representation of a scalar two-loop double box diagram. 
Then, in Sec~\ref{sec:numeval}, we discuss the numerical evaluation of 
causal integrands and present the routines available in \Lotty{} to numerically evaluate them.
We remark the integration procedures carried out in Ref.~\cite{Aguilera-Verdugo:2020nrp} 
and provide the numerical integration of three-point 
planar and non-planar scalar triangles at two loops. 
To check our results, we use the available codes that have implemented
the sector decomposition algorithm~\cite{Hepp:1966eg,Roth:1996pd,Binoth:2000ps,Heinrich:2008si}, 
\SecDec~\cite{Borowka:2015mxa} 
and \Fiesta~\cite{Smirnov:2015mct}. 
Finally, in Sec.~\ref{sec:conclusions},
we draw our conclusions and summarise the features of \Lotty{}.

\section{Loop-tree duality}
\label{sec:ltd0}

In this section, we set the notation for the multi-loop Feynman integrals and recap the 
main features of the novel formulation of the loop-tree duality (LTD) developed in Ref.~\cite{Verdugo:2020kzh}. 

In order to start the discussion, let us consider
a generic $N$-point integral at $L$ loops in the Feynman representation, 
\begin{align}
\mathcal{A}_{N}^{\left(L\right)}\left(1,\hdots,r\right) & =\int_{\ell_{1},\hdots,\ell_{L}}d\mathcal{A}_{N}^{\left(L\right)}\left(1,\hdots,r\right)\,.
\label{eq:myintL}
\end{align}
Here and in the following, to simplify the notation in the integration measure, we define
$\int_{\ell_{s}}\equiv-\imath\mu^{4-d}\int d^{d}\ell_{s}/\left(2\pi\right)^{d}$ and, 
\begin{align}
d\mathcal{A}_{N}^{\left(L\right)}\left(1,\hdots,r\right)&=  \mathcal{N}\left(\left\{ \ell_{i}\right\} _{L},\left\{ p_{j}\right\} _{N}\right)\times 
G_{F}\left(1^{\alpha_{1}},\hdots,r^{\alpha_{r}}\right)\,.
%G_{F}\left(1,\hdots,r\right)\,.
\label{eq:int}
\end{align}
Thus, the integral in Eq.~\eqref{eq:myintL} is understood as 
 an $L$-loop topology with $r$ internal lines.
Since each internal line is characterised by a propagator of the form
$q_{i_r}=\ell_r+k_{i_r}$ (with $i_r\in\{1,2,\hdots,r\}$), 
where $\ell_r$ is the loop momentum identifying
this set, and $k_{i_r}$ accounts for a linear combination of external momenta $\{p_j\}_N$,
one can group lines that share the same integration momenta.
For instance, one-, two- and three-loop scattering amplitudes, 
as shall be described in Sec.~\ref{sec:refinedual}, are characterised 
by one, three and six combinations of loop momenta, respectively. 
The numerator $\mathcal{N}$ is a function of the loop and external momenta, whose structure 
is given by the Feynman rules of the theory. 
The function $G_F$ collects all the Feynman propagators of the topology,
\begin{align}
G_{F}\left(1^{\alpha_{1}},\hdots,r^{\alpha_{r}}\right)=\prod_{i\in1\cup2\cdots\cup r}\left(G_{F}\left(q_{i}\right)\right)^{\alpha_{i}}\,,
\label{eq:gfall}
\end{align}
where $\alpha_{i}$ are the powers of the propagators and, 
\begin{align}
G_{F}\left(q_{i}\right) & =\frac{1}{q_{i}^{2}-m_{i}^{2}+\imath0}=\frac{1}{\left(q_{i,0}+q_{i,0}^{\left(+\right)}\right)
\left(q_{i,0}-q_{i,0}^{\left(+\right)}\right)}\,,
\label{eq:gf}
\end{align}
the usual Feynman propagator of a one single particle, with $m_{i}$
its mass, $+\imath0$ the infinitesimal imaginary Feynman prescription
and, 
\begin{align}
q_{i,0}^{\left(+\right)} & =\sqrt{\boldsymbol{q}_{i}^{2}+m_{i}^{2}-\imath0}\,,
\end{align}
the on-shell energy of the loop momentum $q_{i}$ written in terms
of the spatial components~$\boldsymbol{q}_{i}$. 

Let us note that in the definition of Feynman propagators~\eqref{eq:gf},
we explicitly pulled out the dependence on the energy component of the loop momenta, $q_{i,0}$.
The latter is carried out in order to profit of the Cauchy residue theorem 
and integrate out one degree of freedom, which corresponds to the energy
component of each loop momentum. 
Thus, to compute the residues, one selects the poles with negative imaginary part in the complex
plane of the energy component of the loop momentum that is integrated,
as explained in Ref.~\cite{Verdugo:2020kzh}.

In view of the representation of the Feynman propagators, in terms of the energy component 
of the loop momenta, the numerator of Eq.~\eqref{eq:int} can also be expressed as polynomial in
these variables, 
\begin{align}
\mathcal{N}=&\sum_{i=1}^{R}c_{i}\,\prod_{j=1}^{L}\,\ell_{j,0}^{\alpha_{ij}}\,,
\label{eq:numq0}
\end{align}
where $c$ are coefficients in terms of 
the kinematic variables and the 
spatial components of the loop momenta, $\alpha_{ij}$ are positive integer numbers.
This decomposition can be obtained from the $d$-dimensional scalar product,
\begin{align}
q_{i}\cdot q_{j}&=q_{i}^{\left(+\right)}\cdot q_{j}^{\left(+\right)}-q_{i,0}^{\left(+\right)}q_{j,0}^{\left(+\right)}+q_{i,0}q_{j,0}\,,
\end{align}
where $q_{i}^{(+)}$ corresponds to the $d$-dimensional momentum with on-shell energy $\qpw{i}$. 

Hence, the dual representation of the integrand (\ref{eq:int}), after
setting on shell the propagators that depend on the loop momentum
$q_{i_{1}}$, is defined as follows, 
\begin{align}
\mathcal{A}_{D}^{\left(L\right)}\left(1,\hdots,r\right) & \equiv\sum_{i_{1}\in1}\text{Res}\left(d\mathcal{A}_{N}^{\left(L\right)}\left(1,\hdots,r\right),\text{Im}\left(q_{i_{1},0}\right)<0\right)\,,
\end{align}
where the factor $-2\pi\imath$ that comes from the Cauchy residue
theorem is absorbed in the definition of the integration measure as
shall be noted below. As mentioned before, this residue corresponds to integrating
out the energy components of the loop momenta and allows to introduce the 
nested residue as follows, 
\begin{align}
\mathcal{A}_{D}^{\left(L\right)}\left(1,\hdots,s;s+1,\hdots,r\right) & \equiv\sum_{i_{s}\in s}\text{Res}\left(\mathcal{A}_{D}^{\left(L\right)}\left(1,\hdots,s-1;s,\hdots,r\right),\text{Im}\left(q_{i_{s},0}\right)<0\right)\,,
\label{eq:nestres}
\end{align}
where the iteration goes until the $s$-th set and corresponds to
setting simultaneously $L$ lines on shell. The latter is equivalent
to open the loop topology (or amplitude) into connected trees. 

Finally, with the integration of the energy component of the loop
momentum, one passes from Minkowski to Euclidean space. In the following,
we use the abbreviation, 
\begin{align}
\int_{\vec{\ell}_{s}}\bullet\equiv & -\mu^{d-4}\int\frac{d^{d-1}\ell_{s}}{\left(2\pi\right)^{d-1}}\bullet\,,
\label{eq:defl1}
\end{align}
for the $\left(d-1\right)$-momentum integration measure.

\section{Loop-tree duality in \Lotty}
\label{sec:ltd1}

\Lotty{} -- The LOop-Tree dualiTY automation, is a standalone \Mathematica{} package 
focused on the evaluation of scattering amplitudes by following the ideas of 
the novel formulation of the loop-tree duality formalism. 
The routines implemented in \Lotty{} apply, straightforwardly and systematically,
the ideas developed in Refs.~\cite{Verdugo:2020kzh,Ramirez-Uribe:2020hes,Aguilera-Verdugo:2020nrp,Aguilera-Verdugo:2020kzc,Bobadilla:2021rmu}. 
To illustrate the implementation of LTD described in Sec.~\ref{sec:ltd0}, 
we present, in this section, the main functions of \Lotty{} to find 
the dual representation of multi-loop Feynman integrands
together with the decomposition of scattering amplitudes at two and three loops.

\subsection{Installation}

In order to use \Lotty{}, the user needs to have \Mathematica{} installed (version 10 or higher).
This package can be downloaded from the public Bitbucket repository, 
\begin{equation*}
  \textrm{\href{https://bitbucket.org/wjtorresb/lotty/src/master/}{\color{blue}git clone https://wjtorresb@bitbucket.org/wjtorresb/lotty.git}}
\end{equation*}
Then, to make use of all the features of \Lotty{}, especially for the parallelisation, 
we recommend defining the variable \verb"$LottyPath" with the path to \Lotty{}. 
This can be done by specifying this path in the \texttt{init.m} file of \Mathematica{}.
A suggestion for the latter is, 
{\small
\begin{mmaCell}{Print}
$LottyPath = "/Users/torres/Documents/Repos/lotty";
If[Not[MemberQ[$Path,$LottyPath]],$Path = Flatten[\{$Path, $LottyPath\}]];
\end{mmaCell}
}\noindent 
Writing the path to \Lotty{} also allows us to simply load this package with the 
\Mathematica{} command \texttt{Get["Lotty`"]}. 
In the notebook, one may load the package as follows, 
{\small
\begin{mmaCell}[moredefined={$LottyParallel,$LottyKernels,Lotty}]{Input}
$LottyParallel = True;
$LottyKernels  = 4;
<< Lotty`
\end{mmaCell}
}\noindent
Notice that the variables \texttt{\$LottyParallel} and \texttt{\$LottyKernels}
are only needed to parallelise \Lotty{},
otherwise they can be omitted.
In fact, by default \texttt{\$LottyParallel = False} 
and \texttt{\$LottyKernels} is a variable with an unassigned number.
Assuming one is interested in parallelising, the user has 
to provide
the number of kernels on which \Lotty{} is going to be loaded. 
A naive choice of the latter, using the complete functionality of 
the computer, is setting \texttt{\$LottyKernels = 2\$ProcessorCount}.

\subsection{\texttt{ResidueW}}

As briefly recalled in Sec.~\ref{sec:ltd0}, we need to extensively make use of a routine 
that analytically extracts the residue of a rational function. 
Unfortunately, the \Mathematica{} built-in routine \texttt{Residue} 
turns out not to be efficient when the number of variables of a given
rational function starts increasing. 
Hence, to overcome this issue and make the extraction of residues 
more efficient, we provide our own 
version of this routine, \verb"ResidueW", which relies on first principles.
For instance, the residue of the rational function $f(x)$ 
with a pole of order $m$
at the point $x=x_0$ is evaluated as follows,  
\begin{align}
\text{Res}\left(f\left(x\right),x_{0}\right)&=\lim_{x\to x_{0}}\frac{1}{(m-1)!}\frac{d^{m-1}}{dz^{m-1}}\left[\left(x-x_{0}\right)^{m}f\left(x\right)\right]\,.
\label{eq:resw}
\end{align}
Hence, if $f(x)$ has a single pole at the point $x=x_0$, we set $m=1$.

Since the denominator of the rational function is constructed from a product of 
propagators (see Eq.~\eqref{eq:gf}), the extraction of the residue 
is straightforward due to cancellations that occur before taking the limit of Eq.~\eqref{eq:resw}.

Let us illustrate the performance of \verb"Residue" and \verb"ResidueW"
in a simple rational function with three variables, 
{\small
\begin{mmaCell}[moredefined={ResidueW}]{Input}
\mmaFrac{1}{(\mmaSup{y1}{2}-\mmaSup{x}{2})((\mmaSup{y2}{2}-\mmaSup{x}{2})(\mmaSup{(y2-y1)}{2}-\mmaSup{x}{2})}// Residue[#, \{x, y1 - y2\}] &; // AbsoluteTiming

\mmaFrac{1}{(\mmaSup{y1}{2}-\mmaSup{x}{2})((\mmaSup{y2}{2}-\mmaSup{x}{2})(\mmaSup{(y2-y1)}{2}-\mmaSup{x}{2})}// ResidueW[#, \{x, y1 - y2\}] &; // AbsoluteTiming
\end{mmaCell}
}
{\small
\begin{mmaCell}{Output}
{0.04661, Null}
\end{mmaCell}
}\vspace{-3ex}
{\small
\begin{mmaCell}{Output}
{0.002796, Null}
\end{mmaCell}
}\noindent 
Clearly, the efficiency of \verb"Residue" w.r.t. \verb"ResidueW" starts
decreasing when more variables are present in the calculation. 
For instance, when considering rational functions in terms of more than 
five variables, e.g. loop topologies with five internal lines, 
the performance of \texttt{Residue} does not allow to go further in the 
evaluation of residues. 
On the contrary, \texttt{ResidueW} overcomes this issue,
making the nested application of the residue~\eqref{eq:nestres} 
more efficient and simpler. 
In all the applications considered so far, dual decompositions of up-to four-loop topologies, 
we have not observed any limitation in the extraction of the residue.

\subsection{\texttt{GetDual}}

Then, with an improved version of \texttt{Residue}, 
we can start describing the various routines that automate the LTD formalism. 
This is the case of \verb"GetDual" that
generates the residue in terms of the on-shell energies, $q_{i,0}^{(+)}$, provided
that the integrand is a rational function in the energy components of the 
loop momenta, as displayed in Eq.~\eqref{eq:numq0}.
Hence, the entries of \verb"GetDual" are the numerator, 
list of propagators and list of loop momenta. 
Notice that the explicit dependence of the masses
for each propagator is not necessary, since they are included in 
the definition of $q_{i,0}^{(+)}$. 

\begin{figure}[t]
\centering
\includegraphics[scale=1]{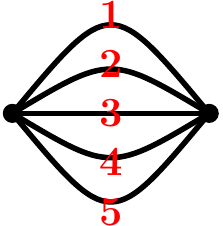}
\qquad
\includegraphics[scale=1]{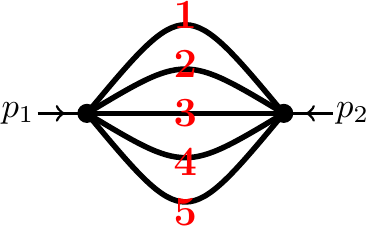}
\caption{Four-loop sunrise diagrams with and without the insertion of external 
momenta in the vertices. }
\label{fig:sun4L}
\end{figure}

To illustrate the use of this function, let us consider 
a scalar sunrise diagram, without external momenta, 
at four loops, as the one depicted in Fig.~\ref{fig:sun4L}~(left),
{\small
\begin{mmaCell}[moredefined={numerator,propagators,loopmom}]{Input}
numerator = 1;
propagators = \{l1, l2, l3, l4, - l1 - l2 - l3 - l4\};
loopmom = \{l1, l2, l3, l4\};
\end{mmaCell}
}\vspace{-3ex}
{\small
\begin{mmaCell}[moredefined={tmp,GetDual,numerator, propagators, loopmom}]{Input}
tmp = GetDual[numerator, propagators, loopmom];
tmp// Total // Together
\end{mmaCell}
}
{\small
\begin{mmaCell}[moredefined={tmp,GetDual,numerator, propagators, loopmom}]{Output}
-\mmaFrac{1}{16 \mmaSubSup{q[1]}{0}{(+)}\mmaSubSup{q[2]}{0}{(+)}\mmaSubSup{q[3]}{0}{(+)}\mmaSubSup{q[4]}{0}{(+)}\mmaSubSup{q[5]}{0}{(+)}(\mmaSubSup{q[1]}{0}{(+)}+\mmaSubSup{q[2]}{0}{(+)}+\mmaSubSup{q[3]}{0}{(+)}+\mmaSubSup{q[4]}{0}{(+)}+\mmaSubSup{q[5]}{0}{(+)})}
\end{mmaCell}
}\noindent 
The latter, in the convention of Ref.~\cite{Verdugo:2020kzh}, corresponds to 
a Maximal loop topology (MLT) configuration at
four loops with no insertion of external momenta in the vertices. 
In general, the MLT configuration is characterised by $L+1$ subsets of propagators of the form, 
\begin{align}
q_{i_{s}} & =\ell_{i}+p_{i_{s}}\,,\quad(i=1,\hdots,L)\,,\nonumber \\
q_{{L+1}_{s}} & =-\sum_{i=1}^{L}\ell_{i}-p_{{L+1}_{s}}\,,
\label{eq:mlt}
\end{align}
where $s$ accounts for the set of all propagators. 
A further discussion regarding the structure of a loop topology 
in terms of sets of propagators is provided in Sec.~\ref{sec:refinedual}. 
Thus, the particular cases of Fig.~\ref{fig:sun4L} corresponds to $L=4$ and considering 
one element per subset.

Let us notice that $q_{i,0}^{(+)}$ have been labelled  according to their position 
in the list of \texttt{propagators}, e.g. the fifth propagator is internally understood as, 
\begin{align}
q_{5,0}^{(+)}=\sqrt{\left(\boldsymbol{\ell}_{1}+\boldsymbol{\ell}_{2}+\boldsymbol{\ell}_{3}+\boldsymbol{\ell}_{4}\right)^{2}+m_{5}^{2}-\imath0}\,,
\end{align}
where the dependence on the mass and  spatial components of the 
loop momenta is recovered when expanding the on-shell  energies out.

In the former example, we consider,
for simplicity, the numerator $\mathcal{N}=1$,
however, the extension to 
numerators expressed as polynomials in the energy component of the loop momenta,
as displayed in Eq.~\eqref{eq:numq0},
can also be implemented in \verb"GetDual".
The latter is carried out because the LTD formalism works at the level of energies only. 
Then, to implement an arbitrary numerator with the explicit 
dependence on the energy component of the loop momenta, 
we use the \Mathematica{} function \verb"Subscript[#,0]".
For instance, the numerators $\ell_{1,0}$ and $\ell_{1,0}^2$ 
are simply expressed as \texttt{Subscript[l1,0]} and \verb"Subscript[l1,0]^2",
respectively. 
In more details, let us add these numerators to the loop topology 
described above, 
{\small
\begin{mmaCell}[moredefined={tmp,GetDual,numerator, propagators, loopmom}]{Input}
numerator = Subscript[l1, 0];
tmp = GetDual[numerator, propagators, loopmom];
tmp// Total // Together
\end{mmaCell}
}\vspace{-3ex}
{\small
\begin{mmaCell}[moredefined={tmp,GetDual,numerator, propagators, loopmom}]{Output}
0
\end{mmaCell}
}
\noindent 
and, 
{\small
\begin{mmaCell}[moredefined={tmp,GetDual,numerator, propagators, loopmom}]{Input}
numerator = \mmaSup{Subscript[l1, 0]}{2};
tmp = GetDual[numerator, propagators, loopmom];
tmp// Total // Together
\end{mmaCell}
}
{\small
\begin{mmaCell}[moredefined={tmp,GetDual,numerator, propagators, loopmom}]{Output}
\mmaFrac{\mmaSubSup{q[2]}{0}{(+)}+\mmaSubSup{q[3]}{0}{(+)}+\mmaSubSup{q[4]}{0}{(+)}+\mmaSubSup{q[5]}{0}{(+)}}{16 \mmaSubSup{q[2]}{0}{(+)}\mmaSubSup{q[3]}{0}{(+)}\mmaSubSup{q[4]}{0}{(+)}\mmaSubSup{q[5]}{0}{(+)}(\mmaSubSup{q[1]}{0}{(+)}+\mmaSubSup{q[2]}{0}{(+)}+\mmaSubSup{q[3]}{0}{(+)}+\mmaSubSup{q[4]}{0}{(+)}+\mmaSubSup{q[5]}{0}{(+)})}
\end{mmaCell}
}
\noindent 
Clearly, the former is zero because of the absence of external momenta.
The latter, on the contrary, can be understood as, 
\begin{align}
\int_{\ell_{1},\hdots,\ell_{4}}\ell_{1,0}^{2}\,G_{F}\left(1,2,3,4,5\right)=\int_{\ell_{1},\hdots,\ell_{4}}\left(G_{F}\left(2,3,4,5\right)+\left(q_{1,0}^{\left(+\right)}\right)^{2}G_{F}\left(1,2,3,4,5\right)\right)\,.
\end{align}

Let us now consider loop topologies with the insertion of external momenta. 
As mentioned in Sec.~\ref{sec:ltd0}, the LTD formalism performs the extraction of 
residues  by selecting the poles
with negative imaginary part in the complex plane of the energy component
of the loop momentum that is integrated, $\text{Im}\left(q_{i,0}\right)<0$.
Hence, when including external momenta, further assumptions
have to be added to account for the latter. 
Thus, if the loop topology contains external momenta, say $p_i$,
one adds for each momentum the assumption $\text{Im}\left(p_{i,0}\right)=0$
in the flag \texttt{"Assumptions"} of the routine \texttt{GetDual}. 
To illustrate this notation, 
let us consider the same four-loop topology, described above, with the presence of one 
independent external momentum $p_1$, (see Fig.~\ref{fig:sun4L}~(right)), 
{\small
\begin{mmaCell}[moredefined={numerator,propagators,loopmom,assumptions,tmp,GetDual}]{Input}
numerator = 1;
propagators = \{l1, l2, l3, l4, - l1 - l2 - l3 - l4 - p[1]\};
loopmom = \{l1, l2, l3, l4\};
assumptions = \{Im[Subscript[p[1], 0]] == 0\};
tmp = GetDual[numerator, propagators, loopmom,"Assumptions"->assumptions];
tmp// Total // Together// Apart[#, Subscript[p[1], 0]] &
\end{mmaCell}
}
{\small
\begin{mmaCell}[]{Output}
-\mmaFrac{1}{32 \mmaSubSup{q[1]}{0}{(+)}\mmaSubSup{q[2]}{0}{(+)}\mmaSubSup{q[3]}{0}{(+)}\mmaSubSup{q[4]}{0}{(+)}\mmaSubSup{q[5]}{0}{(+)}(\mmaSubSup{q[1]}{0}{(+)}+\mmaSubSup{q[2]}{0}{(+)}+\mmaSubSup{q[3]}{0}{(+)}+\mmaSubSup{q[4]}{0}{(+)}+\mmaSubSup{q[5]}{0}{(+)}+\mmaSub{p[1]}{0})}
-\mmaFrac{1}{32 \mmaSubSup{q[1]}{0}{(+)}\mmaSubSup{q[2]}{0}{(+)}\mmaSubSup{q[3]}{0}{(+)}\mmaSubSup{q[4]}{0}{(+)}\mmaSubSup{q[5]}{0}{(+)}(\mmaSubSup{q[1]}{0}{(+)}+\mmaSubSup{q[2]}{0}{(+)}+\mmaSubSup{q[3]}{0}{(+)}+\mmaSubSup{q[4]}{0}{(+)}+\mmaSubSup{q[5]}{0}{(+)}-\mmaSub{p[1]}{0})}
\end{mmaCell}
}
\noindent 
where the dependence on the energy component of 
the external momentum $p_{1,0}$ is completely pulled out after applying LTD. 

In the applications considered until now, we have dealt with loop integrands
in which the dependence on the Feynman propagators is linear.
In other words, we have always set $\alpha_i=1$ in Eq.~\eqref{eq:int}.
However, 
it is pertinent to work with raise powers in the propagators due to the several applications
at multi-loop level one finds. 
To this end, let us consider, yet the same topology, the four-loop integral, 
\begin{align}
I_{\alpha_{1}\hdots\alpha_{5}}^{\left(4\right)} &=
\int_{\ell_{1},\hdots,\ell_{4}}G_{F}\left(1^{\alpha_{1}},2^{\alpha_{2}},3^{\alpha_{3}},4^{\alpha_{4}},5^{\alpha_{5}}\right)\,,
\end{align}
where each $\alpha_i$ is the $i$-th power of each propagator.
This can be done in \texttt{GetDual} by including the flag \texttt{"Powers"}. 
For instance, the dual representation of the integrand $G_{F}\left(1^{2},2,3,4,5\right)$
can be generated as follows, 
{\small
\begin{mmaCell}[moredefined={numerator,propagators,loopmom,assumptions,tmp,GetDual}]{Input}
tmp = GetDual[numerator, propagators, loopmom,
	"Assumptions" -> assumptions,"Powers" -> \{2, 1, 1, 1, 1\}];
\end{mmaCell}
}
\noindent 
obtaining, after summing all contributions up and a partial fractioning w.r.t. $p_{1,0}$, 
{\small
\begin{mmaCell}[moredefined={tmp}]{Input}
tmp // Total // Together // ApartSquareFree[#, Subscript[p[1], 0]] &;
\end{mmaCell}
}
\noindent
where, 
\begin{align}
\text{\texttt{tmp}} = \frac{1}{x_{5}q_{1,0}^{\left(+\right)}}\left[\frac{1}{\lambda_{1}^{+}}\left(\frac{1}{\lambda_{1}^{+}}+\frac{1}{q_{1,0}^{\left(+\right)}}\right)+\left(\lambda_{1}^{+}\to\lambda_{1}^{-}\right)\right]\,,
\label{eq:mltpow}
\end{align}
with $\lambda_1^{\pm}=\sum_{i=1}^{5}q_{i,0}^{(+)}\pm p_{1}$. 
In Sec.~\ref{sec:causal}, a definition of $\lambda_{i}^{\pm}$ will be provided
along the lines of the causal representation of multi-loop Feynman integrands. 

Furthermore, we can consider negative powers, which corresponds to dealing with a numerator,
e.g. 
\begin{align}
G_{F}\left(1,2,3,4,5^{-1}\right) =
\left(q_{5,0}^2-\left(q_{5,0}^{(+)}\right)^2\right)G_{F}\left(1,2,3,4\right)\,,
\end{align}
yielding to, 
{\small
\begin{mmaCell}[moredefined={numerator,propagators,loopmom,assumptions,tmp,GetDual}]{Input}
tmp = GetDual[numerator, propagators, loopmom,
	"Assumptions" -> assumptions,"Powers" -> \{1, 1, 1, 1, -1\}];
tmp// Total // ExpandNumerator // FullSimplify
\end{mmaCell}
}
{\small
\begin{mmaCell}[]{Output}
\mmaFrac{\mmaSup{(\mmaSubSup{q[1]}{0}{(+)}+\mmaSubSup{q[2]}{0}{(+)}+\mmaSubSup{q[3]}{0}{(+)}+\mmaSubSup{q[4]}{0}{(+)}+\mmaSub{p[1]}{0})}{2}-\mmaSup{(\mmaSubSup{q[5]}{0}{(+)})}{2}}{16 \mmaSubSup{q[1]}{0}{(+)}\mmaSubSup{q[2]}{0}{(+)}\mmaSubSup{q[3]}{0}{(+)}\mmaSubSup{q[4]}{0}{(+)}}
\end{mmaCell}
}
\smallskip
Let us recall that in the applications considered so far, with and without 
insertion of external momenta, we have summed over all ``dual''
integrands through the built-in \Mathematica{} routines \verb"Total" and \verb"Together".
The explicit structure of individual residues coming from the application of LTD
shall be elucidated in the following sections. 
Besides, these routines in \Mathematica{} become inefficient 
when increasing the loop order together with the number of internal lines. 
Hence, as shall be presented in Sec.~\ref{sec:causal}, a representation for 
the sum of these integrands is given, allowing for straightforward numerical evaluation.

\subsection{\texttt{RefineDual}}
\label{sec:refinedual}

In the previous section, we provide the dual representation of a four-loop
sunrise diagram by the direct application of LTD
in terms of the on-shell energies~$q_{i,0}^{(+)}$.
However, it would be beneficial for the generation of a dual 
scattering amplitude to have beforehand a recipe to extract the needed residues
of the latter at a given loop order. 
This study has extensively carried out for topologies up-to
three~\cite{Verdugo:2020kzh,Aguilera-Verdugo:2020nrp} 
and four~\cite{Ramirez-Uribe:2020hes} loops. 
In this section, we elaborate on these results and show
how, within \Lotty{} framework, the parametric form of
these residues, regardless of the loop order, can be provided. 
To this end, we provide the routine \texttt{RefineDual}
that performs a 
``reverse engineering'' to find the dual integrands 
in terms of the various $q_{i,0}$. 

In order to illustrate the use of \texttt{RefineDual}, let us first recall the LTD formalism at one loop, 
\begin{align}
\int_{\ell_{1}}d\mathcal{A}_{N}^{\left(1\right)}\left(\ell_{1},\{p_j\}_{N}\right) & =
\int_{\boldsymbol{\ell}_{1}}G_{D}(q_{1}) = 
\int_{\boldsymbol{\ell}_{1}}\sum_{i=1}^{N}G_{D}\left(i\right)\,,
\end{align}
where the sum runs over all $N$ internal propagators, 
which are parametrically written as $q_{1_s}=\ell_1+k_s$. 
The latter, in effect, corresponds to considering one set
of loop momenta, $\{q_{1}\}$, with $N$ elements. 
We define the dual contributions as, 
\begin{align}
G_{D}\left(\pm i\right)&
\equiv\pm\text{Res}\left(d\mathcal{A}_{N}^{\left(1\right)},\left\{q_{i,0}=\pm q_{i,0}^{\left(+\right)}\right\} \right)\,,
\label{eq:GD1}
\end{align}
where the $-2\pi\imath$ factor has been absorbed in the integration measure
$\int_{\boldsymbol{\ell}_{1}}$, previously defined in Eq.~\eqref{eq:defl1}. 
The minus sign in $-i$ indicates a reversal of the momentum flow of the corresponding propagators.
Notice that this equation is valid independently of the structure of the numerator,
provided that the numerator is polynomial in $q_{i,0}$.
With this in mind, let us recap the most general decomposition of two- and three-loop 
Feynman integrals in terms of dual contributions, \verb"GD". 
\begin{enumerate}[i.]
\item\label{item:2L} 
The most general two-loop Feynman integral is characterised by the set of propagators, 
$\left\{ q_{1_{s}},q_{2_{s}},q_{3_{s}}\right\} $ with $q_{1_{s}},q_{2_{s}}$ and $q_{3_{s}}$
subsets of internal propagators of the form,
\begin{align}
 & q_{1_{s}}=\ell_{1}+p_{1_{s}}\,, &  & q_{2_{s}}=\ell_{2}+p_{2_{s}}\,, &  & q_{3_{s}}=-\ell_{1}-\ell_{2}-p_{3_{s}}\,.
 \label{eq:mlt2}
\end{align}
Then, due to the way how LTD is performed at two loops, two loop momenta are simultaneously 
set on-shell.
Therefore, we can consider, for the sake of simplicity, the subsets $q_{i_s}$ with only one element.
Thus, by applying LTD through \texttt{GetDual},\footnote{Here and in the subsequent case at three loops,
we set, without the loss of generality, external momenta to zero, $p_i=0$. 
Cases at two and three loops with insertion of external momenta
are considered in Secs.~\ref{sec:db_res} and~\ref{sec:tennis_res},
recovering the same decomposition presented in this section.}
{\small
\begin{mmaCell}[moredefined={numerator,propagators,loopmom}]{Input}
numerator = 1;
propagators = \{l1, l2, - l1 - l2\};
loopmom = \{l1, l2\};
\end{mmaCell}
}\vspace{-2ex}
{\small
\begin{mmaCell}[moredefined={tmp,tmp1,GetDual,numerator, propagators, loopmom,RefineDual}]{Input}
tmp  = GetDual[numerator, propagators, loopmom];
tmp1 = RefineDual[tmp, numerator, propagators, loopmom]
\end{mmaCell}
}\vspace{-2ex}
{\small
\begin{mmaCell}[]{Output}
\{GD[\{-1, -2\}], GD[\{2, -3\}], GD[\{-1, -3\}]\}
\end{mmaCell}
}
\noindent
where \texttt{GD[{i,j}]} correspond to the nested application of the
residue~\eqref{eq:GD1},
and are obtained for the specific ordering of the loop momenta $\{\ell_1,\ell_2\}$. 
On the contrary, applying the Cauchy residue theorem on the integrand 
in $\{\ell_2,\ell_1\}$ yields to, 
{\small
\begin{mmaCell}[moredefined={tmp,tmp1,GetDual,numerator, propagators, loopmom,RefineDual}]{Input}
tmp  = GetDual[numerator, propagators, Reverse@loopmom];
tmp1 = RefineDual[tmp, numerator, propagators, Reverse@loopmom]
\end{mmaCell}
}%\vspace{-2ex}
{\small
\begin{mmaCell}[]{Output}
\{GD[\{-1, -2\}], GD[\{1, -3\}], GD[\{-2, -3\}]\}
\end{mmaCell}
}\noindent
that corresponds to the exchange of $\ell_1\leftrightarrow\ell_2$ w.r.t. the former evaluation. 
Notably, as shall be discussed in Sec.~\ref{sec:causal}, 
the ordering in which the residues are computed does not alter
the final result, although individual terms are modified. 
For an extensive discussion on this subject, we refer the reader to Ref.~\cite{Aguilera-Verdugo:2020kzc}. 

Although this result is for the particular case of 
a two-loop topology with three internal lines,
it can be generalised to the general case with three subsets,
as proven in Ref.~\cite{Aguilera-Verdugo:2020nrp}. 
We elucidate this transition in Sec.~\ref{sec:db_res} 
for the two-loop double box topology. 
A pictorial representation of the lines $i$ is displayed in Fig.~\ref{fig:2Ltopos}~(left).
\begin{figure}[t]
\centering
\includegraphics[scale=1]{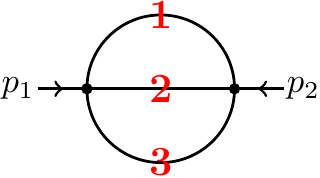}
\qquad
\includegraphics[scale=1]{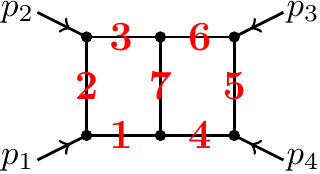}
\caption{Examples of Maximal loop topologies at two loops, the sunrise (left) and the double-box (right) topologies.
The red label corresponds to the propagator number.}
\label{fig:2Ltopos}
\end{figure}
\begin{figure}[t]
\centering
\includegraphics[scale=1.3]{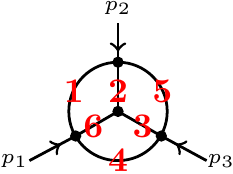}
\qquad
\includegraphics[scale=1]{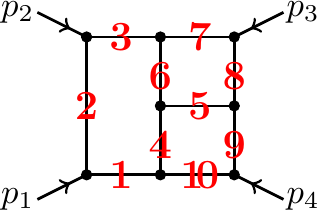}
\caption{Examples of Next-to-Next-to Maximal loop topologies at three loops, 
Mercedes-Benz (left) and tennis-court (right) topologies.
The red label corresponds to the propagator number.}
\label{fig:3Ltopos}
\end{figure}
\item\label{item:3L} 
Similarly, for the most general three-loop Feynman integrals, we have to deal with six subsets, 
which are pictorially displayed in Fig.~\ref{fig:3Ltopos}~(left), 
\begin{align}
q_{i_{s}} & =\ell_{i}+p_{i_{s}}\,,\quad(i=1,2,3)\,,\nonumber \\
q_{4_{s}} & =-\ell_{1}-\ell_{2}-\ell_{3}-p_{4_{s}}\,,\nonumber \\
q_{5_{s}} & =-\ell_{1}-\ell_{2}-p_{5_{s}}\,,\nonumber \\
q_{6_{s}} & =-\ell_{2}-\ell_{3}-p_{6_{s}}\,,
\label{eq:n2mlt3}
\end{align}
whose implementation in \Lotty{} is, 
{\small
\begin{mmaCell}[moredefined={numerator,propagators,loopmom}]{Input}
numerator = 1;
propagators = \{l1, l2, l3, -l1 - l2 - l3, -l1 - l2, -l2 - l3\};
loopmom = \{l1, l2, l3\};
\end{mmaCell}
}
{\small
\begin{mmaCell}[moredefined={tmp,tmp1,GetDual,numerator, propagators, loopmom,RefineDual}]{Input}
tmp  = GetDual[numerator, propagators, loopmom];
tmp1 = RefineDual[tmp, numerator, propagators, loopmom]
\end{mmaCell}
}
{\small
\begin{mmaCell}[]{Output}
\{ GD[\{-1,-2,-3\}],GD[\{-2,-3,4\}],GD[\{-2,-3,5\}],GD[\{1,-3,4\}],
  GD[\{1,-3,5\}],GD[\{-1,-3,6\}],GD[\{3,-4,-6\}],GD[\{3,-5,-6\}],
  GD[\{-1,-2,-4\}],GD[\{2,-4,5\}],GD[\{-1,-4,5\}],GD[\{4,-5,6\}],
  GD[\{1,-2,-6\}],GD[\{-2,-4,-6\}],GD[\{-2,-5,-6\}],GD[\{-1,-5,6\}] \}
\end{mmaCell}
}
\noindent 
The use of this function for any three-loop topology shall be elucidated in Sec.~\ref{sec:tennis_res},
by considering the three-loop tennis-court diagram. 
\end{enumerate}
The aforementioned topologies, in the convention of Ref~\cite{Verdugo:2020kzh}, correspond to MLT
and Next-to-Next-to Maximal loop topology (N$^2$MLT) configurations 
at two and three loops, respectively.
Similar to the definition of MLT summarised in Eq.~\eqref{eq:mlt}, 
let us briefly recall that the N$^2$MLT configuration is characterised by $L+3$  subsets 
of propagators of the form, 
\begin{align}
q_{i_{s}} & =\ell_{i}+p_{i_{s}}\,,\quad(i=1,\hdots,L)\,,\nonumber \\
q_{{L+1}_{s}} & =-\sum_{i=1}^{L}\ell_{i}-p_{{L+1}_{s}}\,,\nonumber \\
q_{{L+2}_{s}} & =-\ell_{1}-\ell_{2}-p_{{L+2}_{s}}\,,\nonumber \\
q_{{L+3}_{s}} & =-\ell_{2}-\ell_{3}-p_{{L+3}_{s}}\,.
\end{align}
whose dual decomposition, regardless of the loop order, 
has been studied by means of convolution and factorisation identities. 

In the applications of LTD studied in Refs.~\cite{Verdugo:2020kzh,Ramirez-Uribe:2020hes,Aguilera-Verdugo:2020nrp},
we were interested in achieving a decomposition of topologies of up-to 
four loops in terms of nested residues. 
The motivation to do so is to have a parametric representation,
regardless of the loop order, in terms of the latter. 
Thus, allowing to provide the  dual representation of any multi-loop
scattering amplitude by only applying the corresponding residues. 
For instance, computing the dual representation of 
a three-loop scattering amplitude corresponds only
to evaluating the residues \texttt{GD[\{i,j,k\}]} obtained 
in item~\ref{item:3L}. 
We remark that using the routine \texttt{RefineDual} 
allowed us to cross-check our conjectures and inspired us
to provide mathematical proofs for the latter. 

\subsection{Two-loop double box}
\label{sec:db_res}
In the former section, it was discussed that the dual decomposition of any
two-loop scattering amplitude can be obtained from the evaluation
of a two-loop vacuum diagram (see  Fig.~\ref{fig:2Ltopos}~(left)),
where its internal lines are promoted to be sets of propagators.
Hence, to elucidate the connection between the most general 
loop topology at two loops and the one obtained for a specify 
two-loop diagram, 
say two-loop double box, we evaluate the latter as follows, 
{\small
\begin{mmaCell}[moredefined={numerator,propagators,loopmom,assumptions}]{Input}
numerator = 1;
propagators = \{l1, l1+p[1], l1+p[1]+p[2], l2, l2+p[4], l2+p[3]+p[4], -l1-l2\};
loopmom = \{l1, l2\};
assumptions = \{Im[Subscript[p[1], 0]] == 0, Im[Subscript[p[2], 0]] == 0, 
		Im[Subscript[p[3], 0]] == 0, Im[Subscript[p[4], 0]] == 0\};
\end{mmaCell}
}\noindent
Although  we set \texttt{numerator = 1}, the very same analysis can be 
carried out for any arbitrary polynomial in the energy components of the 
loop momenta. 
Thus, evaluating all residues, one gets,
{\small
\begin{mmaCell}[moredefined={tmp,tmp1,GetDual,numerator, propagators, loopmom,RefineDual,assumptions}]{Input}
tmp  = GetDual[numerator, propagators, loopmom,"Assumptions" -> assumptions];
tmp1 = RefineDual[tmp, numerator, propagators, loopmom]
\end{mmaCell}
}\vspace{-3ex}
{\small
\begin{mmaCell}[]{Output}
\{ GD[\{1,4\}],GD[\{2,4\}],GD[\{3,4\}],GD[\{4,-7\}],GD[\{1,5\}],GD[\{2,5\}],
  GD[\{3,5\}],GD[\{5,-7\}],GD[\{1,6\}],GD[\{2,6\}],GD[\{3,6\}],GD[\{6,-7\}],
  GD[\{-1,-7\}],GD[\{-2,-7\}],GD[\{-3,-7\}] \}
\end{mmaCell}
}\noindent
Notice that from the list of \texttt{propagators} 
one can distinguish three subsets according to the dependence 
on the loop momenta,
\begin{align}
\left\{\alpha_1,\alpha_2,\alpha_3\right\} &= 
\left\{ \left\{ \ell_{1},\ell_{1}+p_{1},\ell_{1}+p_{1}+p_{2}\right\} ,\left\{ \ell_{2},\ell_{2}-p_{4},\ell_{2}-p_{3}-p_{4}\right\} ,\left\{ \ell_{1}+\ell_{2}\right\} \right\} \,,
\end{align}
then, relating each propagator of the double box to its subset $\alpha_i$, 
{ \small
\begin{mmaCell}[moredefined={tmp1}]{Input}
tmp1 /. \{a_ /; MemberQ[\{1, 2, 3\}, Abs[a]] :> Sign[a] \(\alpha\)1, 
	 a_ /; MemberQ[\{4,5,6\}, Abs[a]] :> Sign[a] \(\alpha\)2, 
	 a_ /; MemberQ[\{7\}, Abs[a]] :> Sign[a] \(\alpha\)3 \}
% /. GD[a__] :> GD[Sort@a] // Union	 
\end{mmaCell}
}\vspace{-3ex}
{ \small
\begin{mmaCell}{Output}
\{ GD[\{-\(\alpha\)1, -\(\alpha\)3\}], GD[\{\(\alpha\)1, \(\alpha\)2\}], GD[\{\(\alpha\)2, -\(\alpha\)3\}] \}
\end{mmaCell}
}
\noindent which corresponds to the same structure of the two-loop residues, parametrised 
according to Eq.~\eqref{eq:mlt2} and obtained in item~\ref{item:2L}, 
after trivially changing $\alpha_i\to i$.

\subsection{Three-loop tennis court}
\label{sec:tennis_res}

Similar to the two-loop case, let us discuss a particular example at three loops,
say three-loop tennis-court diagram (see Fig.~\ref{fig:3Ltopos}~(right)), 
and compare it with the residues obtained from the evaluation 
of the most general topology at three loops, listed in item~\ref{item:3L}. 

Since the decomposition at three loops is assumed to be valid for any 
three-loop scattering amplitudes, let us consider, differently from the previous case,
the numerator
\begin{align}
\mathcal{N} & =a_{1}\,\ell_{1,0}^{3}+a_{2}\,\ell_{2,0}\,\ell_{3,0}+a_{3}\,\ell_{3,0}^{2}\,,
\end{align}
with $a_i$ arbitrary constants independent of the on-shell energies. 

Then, in \Lotty{}, we parametrise the tennis-court diagram as, 
{\small
\begin{mmaCell}[moredefined={tmp,GetDual,numerator, assumptions,propagators, LoopMom}]{Input}
LoopMom     = \{l1, l2, l3\};
assumptions = (Im[Subscript[p[#], 0]] == 0) & /@ Range[4];
propagators = \{ l1, l1 + p[1], l1 + p[1] + p[2], l2,
		l3, -l2 - l3, -l1 - l2 - l3 - p[1] - p[2],
		-l1 - l2 - l3 - p[1] - p[2] - p[3], 
		-l1 - l2 + p[4], -l1 - l2 \};
numerator  = ( a[1] \mmaSup{Subscript[l1, 0]}{3} 
		+ a[2] Subscript[l2, 0] Subscript[l3, 0] 
		+ a[3] \mmaSup{Subscript[l3, 0]}{2} );
\end{mmaCell}
}
\noindent
and extract the residues, 
{\small
\begin{mmaCell}[moredefined={tmp,GetDual,numerator, propagators, LoopMom,assumptions}]{Input}
tmp = GetDual[numerator, propagators, LoopMom, "Assumptions" -> assumptions];
\end{mmaCell}
}
\noindent
Based on the organisation of the loop momenta in \texttt{propagators}, we can 
distinguish the six sets that represent this topology, 
\begin{align}
\left\{ \alpha_{1},\hdots,\alpha_{6}\right\} = & \Big\{\left\{ \ell_{1},\ell_{1}+p_{1},\ell_{1}+p_{12}\right\} ,\left\{ \ell_{2}\right\} ,\left\{ \ell_{3}\right\} ,\nonumber \\
 & \left\{ \ell_{123}+p_{12},\ell_{123}+p_{123}\right\} ,\left\{ \ell_{12},\ell_{12}-p_{4}\right\} ,\left\{ \ell_{23}\right\} \Big\}\,,
\end{align}
generating, then, the residues, 
{\small
\begin{mmaCell}[moredefined={RefineDual,tmp,numerator,tmp1,propagators,LoopMom}]{Input}
tmp1 = RefineDual[tmp, numerator, propagators, LoopMom]
\end{mmaCell}
}\vspace{-3ex}
{\small
\begin{mmaCell}{Output}
\{ GD[\{1, 4, 5\}], GD[\{2, 4, 5\}], GD[\{3, 4, 5\}], GD[\{4, 5, -7\}], 
  GD[\{4, 5, -8\}], GD[\{4, 5, -9\}], GD[\{4, 5, -10\}], GD[\{1, 5, -6\}], 
  GD[\{2, 5, -6\}], GD[\{3, 5, -6\}], GD[\{5, -6, -7\}], GD[\{5, -6, -8\}], 
  GD[\{5, -6, -9\}], GD[\{5, -6, -10\}], GD[\{-1, 5, -7\}], GD[\{-2, 5, -7\}], 
  GD[\{-3, 5, -7\}], GD[\{-1, 5, -8\}], GD[\{-2, 5, -8\}], GD[\{-3, 5, -8\}], 
  GD[\{-1, 5, -9\}], GD[\{-2, 5, -9\}], GD[\{-3, 5, -9\}], GD[\{-1, 5, -10\}], 
  GD[\{-2, 5, -10\}], GD[\{-3, 5, -10\}], GD[\{1, -4, -6\}], GD[\{2, -4, -6\}],
  GD[\{3, -4, -6\}], GD[\{-4, -6, -7\}], GD[\{-4, -6, -8\}], GD[\{-4, -6, -9\}], 
  GD[\{-4, -6, -10\}], GD[\{-1, -4, -7\}], GD[\{-2, -4, -7\}], GD[\{-3, -4, -7\}], 
  GD[\{-1, -4, -8\}], GD[\{-2, -4, -8\}], GD[\{-3, -4, -8\}], GD[\{2, -6, 9\}], 
  GD[\{3, -6, 9\}], GD[\{4, -7, 9\}], GD[\{-6, -7, 9\}], GD[\{-1, -7, 9\}], 
  GD[\{-2, -7, 9\}], GD[\{-3, -7, 9\}], GD[\{4, -8, 9\}], GD[\{-6, -8, 9\}], 
  GD[\{-1, -8, 9\}], GD[\{-2, -8, 9\}], GD[\{-3, -8, 9\}], GD[\{1, -6, 9\}], 
  GD[\{2, -6, 10\}], GD[\{3, -6, 10\}], GD[\{4, -7, 10\}], GD[\{-6, -7, 10\}], 
  GD[\{-1, -7, 10\}], GD[\{-2, -7, 10\}], GD[\{-3, -7, 10\}], GD[\{4, -8, 10\}], 
  GD[\{-6, -8, 10\}], GD[\{-1, -8, 10\}], GD[\{-2, -8, 10\}], 
  GD[\{-3, -8, 10\}], GD[\{1, -6, 10\}] \}
\end{mmaCell}
}
\noindent
which can be grouped as follows, 
{\small
\begin{mmaCell}[moredefined={RefineDual,tmp,numerator,tmp1,propagators,LoopMom}]{Input}
tmp1 /. \{ a_ /; MemberQ[\{1, 2, 3\}, Abs[a]] :> Sign[a] \(\alpha\)1,
	  a_ /; MemberQ[\{4\}, Abs[a]] :> Sign[a] \(\alpha\)2,
	  a_ /; MemberQ[\{5\}, Abs[a]] :> Sign[a] \(\alpha\)3,
	  a_ /; MemberQ[\{7, 8\}, Abs[a]] :> Sign[a] \(\alpha\)4,
	  a_ /; MemberQ[\{9, 10\}, Abs[a]] :> Sign[a] \(\alpha\)5,
	  a_ /; MemberQ[\{6\}, Abs[a]] :> Sign[a] \(\alpha\)6 \} 
% /. GD[a__] :> GD[Sort@a] // Union
\end{mmaCell}
}\vspace{-3ex}
{ \small
\begin{mmaCell}{Output}
\{ GD[\{-\(\alpha\)1, -\(\alpha\)2, -\(\alpha\)4\}], GD[\{-\(\alpha\)1, \(\alpha\)3, -\(\alpha\)4\}], GD[\{-\(\alpha\)1, \(\alpha\)3, -\(\alpha\)5\}], 
  GD[\{-\(\alpha\)1, -\(\alpha\)4, \(\alpha\)5\}],  GD[\{\(\alpha\)1, -\(\alpha\)2, -\(\alpha\)6\}], GD[\{\(\alpha\)1, \(\alpha\)2, \(\alpha\)3\}], 
  GD[\{\(\alpha\)1, \(\alpha\)3, -\(\alpha\)6\}],   GD[\{\(\alpha\)1, \(\alpha\)5, -\(\alpha\)6\}],  GD[\{-\(\alpha\)2, -\(\alpha\)4, -\(\alpha\)6\}], 
  GD[\{-\(\alpha\)2, -\(\alpha\)5, -\(\alpha\)6\}], GD[\{\(\alpha\)2, \(\alpha\)3, -\(\alpha\)4\}],  GD[\{\(\alpha\)2, \(\alpha\)3, -\(\alpha\)5\}], 
  GD[\{\(\alpha\)2, -\(\alpha\)4, \(\alpha\)5\}],   GD[\{\(\alpha\)3, -\(\alpha\)4, -\(\alpha\)6\}], GD[\{\(\alpha\)3, -\(\alpha\)5, -\(\alpha\)6\}], 
  GD[\{-\(\alpha\)4, \(\alpha\)5, -\(\alpha\)6\}] \}
\end{mmaCell}
}
\noindent
corresponding to the 12 residues obtained for the three-loop vacuum diagram of Fig.~\ref{fig:3Ltopos}~(left) 
and listed in item~\ref{item:3L}.

\section{Causal representation in \Lotty}
\label{sec:causal}

In the previous section, we presented the procedure to generate dual integrands in \Lotty{}
through the application of LTD. 
In the examples considered so far, we have not looked at  the structure of each 
residue in terms of on-shell energies, $q_{i,0}^{(+)}$. 
In fact, we have only been interested in the full sum of these residues. 
This sum can be easily performed in \Mathematica{} for simple topologies, like 
MLT configurations (see Fig.~\ref{fig:sun4L} for the four-loop case), 
since it is always expressed as, 
\begin{align}
\mathcal{A}_{\text{MLT}}^{\left(L\right)} & =\int_{\boldsymbol{\ell}_{1},\hdots,\boldsymbol{\ell}_{L}}\frac{-1}{x_{L+1}}\left(\frac{1}{\lambda_{1}^{+}}+\frac{1}{\lambda_{1}^{-}}\right)\,,
\label{eq:MLTL}
\end{align}
with $x_{L+k}=\prod_{i=1}^{L+k}2q_{i,0}^{\left(+\right)}$ and $\lambda_{1}^{\pm}=\sum_{i=1}^{L+1}q_{i,0}^{\left(+\right)}\pm p_{1,0}$. 
The way in which the integrand is written in Eq.~\eqref{eq:MLTL},
in terms of causal propagators, $\lambda_{i}^{\pm}$, 
corresponds to the causal representation of multi-loop Feynman integrands. 
In the recent work of Ref.~\cite{Bobadilla:2021rmu}, we provide an
interpretation of this pattern by taking into account the features
of loop topologies, cusps and edges.
To start, we discuss that for the MLT configuration 
one has two cusps and one edge,
where an edge corresponds to a set of all
internal lines that connect two cusps,
\begin{align}
\parbox{20mm}{
\begin{tikzpicture}[line width=1.5 pt, node distance=0.8 cm and 0.4 cm]
\coordinate (v1) at (0,0);
\coordinate (v22) at (2,0);
\draw[fermionbar] (v1) -- (v22);
\draw[fill=black] (v1) circle (.05cm);
\draw[fill=black] (v22) circle (.05cm);
\end{tikzpicture}
}
\quad \equiv \quad
\parbox{20mm}{
\begin{tikzpicture}[line width=1 pt, node distance=0.8 cm and 0.4 cm]
\coordinate (v1) at (0,0);
\coordinate (v2) at (1,0);
\coordinate (v22) at (2,0);
\coordinate (v2a) at (1,0.5);
\draw[fermionbar] (v1) -- (v22);
\semiloop[fermion]{v1}{v22}{0};
\semiloop[fermionbar]{v22}{v1}{180};
\draw[fill=black] (v1) circle (.05cm);
\draw[fill=black] (v22) circle (.05cm);
\node  at (1,1.3) {\footnotesize$1$}; 
\node  at (1,0.6) {$\huge\boldsymbol{\vdots}$}; 
\node  at (1,-0.3) {\footnotesize$s$}; 
\node  at (1,-1.3) {\footnotesize$s+1$}; 
\end{tikzpicture}
}\quad \,.
\label{eq:graphrep}
\end{align}
Hence, in the MLT configuration there is only one edge that corresponds
to a set with $L+1$ elements.  
Similarly, the N$^2$MLT configuration can be seen as a topology made of four cusps and six edges,
see Fig.~\ref{fig:3Ltopos}~(left). 
Thus, within this framework, a loop topology made of $k+2$ cusps 
is referred as N$^k$MLT configuration,
whose causal representation can be provided along the lines of Ref.~\cite{Bobadilla:2021rmu}. 
%Due to the simplicity of the MLT configuration, 
%in the following, we elucidate this feature for more involved loop topologies. 

Let us then draw our attention to the  N$^2$MLT configurations, 
Fig.~\ref{fig:3Ltopos}~(left),
with propagators~\eqref{eq:n2mlt3}, where,
to illustrate in more details, we generate the dual integrands 
with external momenta attached to the cusps,
{\small
\begin{mmaCell}[moredefined={numerator,LoopMom,propagators,assumptions}]{Input}
numerator = 1;
LoopMom = \{l1, l2, l3\};
propagators = \{l1, l2, l3, -l1 - l2 - l3 + p[2] + p[3], 
		-l1 - l2 + p[2], -l2 - l3 + p[1] + p[2] + p[3] \};
assumptions = (Im[Subscript[p[#], 0]] == 0) & /@ Range[3];
\end{mmaCell}
}\noindent
Notice that the external momenta satisfy momentum conservation 
$p_1+p_2+p_3+p_4 = 0$. 
Then, the residues are computed as follows, 
{\small
\begin{mmaCell}[moredefined={RefineDual,tmp,tmp1,GetDual,numerator,LoopMom,propagators,assumptions}]{Input}
tmp = GetDual[numerator, propagators, LoopMom, "Assumptions" -> assumptions];
tmp1 = RefineDual[tmp, numerator, propagators, LoopMom];
\end{mmaCell}
}\vspace{-3ex}
{\small
\begin{mmaCell}[]{Output}
\{ GD[\{-1, -2, -3\}], GD[\{-2, -3, 4\}], GD[\{-2, -3, 5\}], GD[\{1, -3, 4\}], 
  GD[\{1, -3, 5\}], GD[\{-1, -3, 6\}], GD[\{3, -4, -6\}], GD[\{3, -5, -6\}], 
  GD[\{-1, -2, -4\}], GD[\{2, -4, 5\}], GD[\{-1, -4, 5\}], GD[\{4, -5, 6\}], 
  GD[\{1, -2, -6\}], GD[\{-2, -4, -6\}], GD[\{-2, -5, -6\}], GD[\{-1, -5, 6\}] \}
\end{mmaCell}
}
\noindent
recovering the same list of residues summarised in item~\ref{item:3L}.

As mentioned above, the explicit expression of individual residues, \texttt{GD}, 
in terms of on-shell energies has not been considered until now.
Hence, let us study, without the loss of generality, 
the explicit structure of \texttt{GD[\{-1,-2,-3\}]} in terms of $\qpw{i}$, 
\begin{align}
\text{\texttt{GD[\{-1,-2,-3\}]}}= & \frac{1}{8q_{1,0}^{\left(+\right)}q_{2,0}^{\left(+\right)}q_{3,0}^{\left(+\right)}}\nonumber \\
 & \times\frac{1}{\left(q_{\left(1,2,5\right),0}^{\left(+\right)}-p_{2,0}\right)\left(q_{\left(1,2,3,4\right),0}^{\left(+\right)}-p_{23,0}\right)\left(q_{\left(2,3,6\right),0}^{\left(+\right)}-p_{123,0}\right)}\nonumber \\
 & \times\frac{1}{\left(q_{\left(1,2,\bar{5}\right),0}^{\left(+\right)}-p_{2,0}\right)\left(q_{\left(1,2,3,\bar{4}\right),0}^{\left(+\right)}-p_{23,0}\right)\left(q_{\left(2,3,\bar{6}\right),0}^{\left(+\right)}-p_{123,0}\right)}\,,
 \label{eq:res_n2mlt}
\end{align}
where we introduce the shorthand notation, 
\begin{align}
& \qpw{(i_1,i_2,\hdots,i_N)}\equiv\sum_{k=i_1,i_2,\hdots,i_N}\qpw{i}\,,
&&
\qpw{\bar{k}}\equiv-\qpw{k}\,.
\label{eq:notation}
\end{align}

In order to study the structure of denominators in Eq.~\eqref{eq:res_n2mlt},
let us recall that on-shell energies are positively defined, $\qpw{i}>0$.
Therefore, the first and second lines in 
the residue~\eqref{eq:res_n2mlt} do not display any unphysical singularity.
The only ones that could occur correspond to causal thresholds, 
given by the values of $p_{i,0}$. 
The third line, on the contrary, displays pseudo-thresholds or unphysical singularities, 
which come from the combination of $\qpw{i}$'s with and without bar, 
according to Eq.~\eqref{eq:notation},  e.g. $\qpw{(1,4,\bar{5})}=\qpw{1}+\qpw{4}-\qpw{7}$.
The same pattern is held for all individual residues.

Nonetheless, it has been studied in Refs.~\cite{Aguilera-Verdugo:2020kzc,Ramirez-Uribe:2020hes} 
that the sum of all residues completely drops the dependence 
of all terms that contain at least a bar, $\qpw{(i_1,\bar{i}_2,\hdots)}$.
The former are called non-causal propagators and will not be taken into account 
in the following discussions. 

\subsection{\texttt{GetCausalProps}}

In view of the structure of individual residues,
it will be desirable, before summing all contributions up, to know the shape of the 
denominators in terms of causal propagators, $\lambda_{i}^{\pm} = \qpw{(i,j,\hdots)}+p_{i,0}$. 
Hence, to generate these causal propagators beforehand, 
we implemented in \Lotty{} the routine \texttt{GetCausalProps}
that scans over all residues and extracts the denominators that 
display the structure of causal propagators. 
To elucidate the use of \texttt{GetCausalProps}, 
let us consider the same N$^2$MLT configuration,
{ \small
\begin{mmaCell}[moredefined={GetCausalProps,tmp,propagators}]{Input}
GetCausalProps[tmp, propagators]
\end{mmaCell}
}\vspace{-3ex}
{ \small
\begin{mmaCell}{Print}
ALL \(\lambda\)[i] ARE STORED IN THE FUNCTION \(\lambda\)2qi0
\end{mmaCell}
}\vspace{-3ex}
{ \small
\begin{mmaCell}{Output}
\{ \mmaSub{p[2]}{0} - \(\lambda\)[1],			 \mmaSub{p[2]}{0} + \(\lambda\)[1],
  \mmaSub{p[3]}{0} - \(\lambda\)[2],			  \mmaSub{p[3]}{0} + \(\lambda\)[2],
  \mmaSub{p[1]}{0} + \mmaSub{p[2]}{0} + \mmaSub{p[3]}{0} - \(\lambda\)[3],	\mmaSub{p[1]}{0} + \mmaSub{p[2]}{0} + \mmaSub{p[3]}{0} + \(\lambda\)[3],
  \mmaSub{p[1]}{0} - \(\lambda\)[4],			  \mmaSub{p[1]}{0} + \(\lambda\)[4],
  \mmaSub{p[2]}{0} + \mmaSub{p[3]}{0} - \(\lambda\)[5],		 \mmaSub{p[2]}{0} + \mmaSub{p[3]}{0} + \(\lambda\)[5],
  \mmaSub{p[1]}{0} + \mmaSub{p[3]}{0} - \(\lambda\)[6],		 \mmaSub{p[1]}{0} + \mmaSub{p[3]}{0} + \(\lambda\)[6],
  \mmaSub{p[1]}{0} + \mmaSub{p[2]}{0} - \(\lambda\)[7],		 \mmaSub{p[1]}{0} + \mmaSub{p[2]}{0} + \(\lambda\)[7] \}
\end{mmaCell}
}\noindent
Clearly, because of momentum conservation, 
\mmaSub{p[1]}{0}+\mmaSub{p[2]}{0}+\mmaSub{p[3]}{0} = -\mmaSub{p[4]}{0},
one recovers the same structure of the causal propagators discussed in Ref.~\cite{Bobadilla:2021rmu}. 
Besides, the definitions of \texttt{$\lambda[i]$}, corresponding to the sums of on-shell energies, 
are stored in the variable \texttt{$\lambda$2qi0}.
Let us remark that the knowledge of  causal propagators, before simplifying the full expression, 
has allowed to generate an Ansatz for the causal representation of  
topologies up-to four loops~\cite{Aguilera-Verdugo:2020kzc,Ramirez-Uribe:2020hes}. 
In fact, by following the approach explained in~\cite{Aguilera-Verdugo:2020kzc},
together with the analytic reconstruction over finite fields,
in the \FiniteFlow{} framework, 
we manage to reconstruct
the causal representation of NMLT and N$^2$MLT. 
Likewise, particular cases of N$^3$MLT and N$^4$MLT 
have also been considered in Ref.~\cite{Ramirez-Uribe:2020hes}. 

In order to provide the causal representation of a given N$^{k}$MLT configuration, 
assuming that this topology only has linear Feynman propagators,\footnote{Notice
that the extension to integrands with raised powers in the propagators 
can yet be provided by performing, at integrand level, the derivative  $\partial/\partial(q_{i,0}^{(+)})^{2}$
along the lines of Ref.~\cite{Aguilera-Verdugo:2020kzc}.}
the integrand in the causal representation 
admits the following decomposition, 
\begin{align}
d\mathcal{A}_{L+k+1}^{\left(L\right)} & =\frac{(-1)^{k+1}}{x_{L+k+1}}\mathcal{F}_{L+k+1}\left(\lambda_{i}^{\pm}\right)\,.
\label{eq:allcausal}
\end{align}
%where an N$^k$MLT configuration is a loop topology made of $k+2$ cusps. 

Thus, to understand the causal representation, we can only focus on $\mathcal{F}$,
since $x_{L+k+1}$ can straightforwardly be reconstructed at any time of the calculation. 
In effect, this prefactor can be obtained with the flag \texttt{"GetPref"} 
of \texttt{GetCausalProps}, 
{ \small
\begin{mmaCell}[moredefined={GetCausalProps,tmp,propagators}]{Input}
GetCausalProps[tmp, propagators,"GetPref" -> True]
\end{mmaCell}
}\vspace{-3ex}
{ \small
\begin{mmaCell}{Output}
64 \mmaSubSup{q[1]}{0}{(+)}\mmaSubSup{q[2]}{0}{(+)}\mmaSubSup{q[3]}{0}{(+)}\mmaSubSup{q[4]}{0}{(+)}\mmaSubSup{q[5]}{0}{(+)}\mmaSubSup{q[6]}{0}{(+)}
\end{mmaCell}
}

\subsection{\texttt{AllCausal}}

From the former discussion, it is clear that, within LTD, it is cumbersome to have a rational 
function that displays the structure of integrands
and allows for an efficient numerical evaluation. 
For this reason, our main aim with this causal representation is to 
find ``causal'' terms that can straightforwardly be evaluated.
On the one hand, there is the subtlety with the unphysical singularities that come from
non-causal propagators and, on the  other hand, is the evaluation time. 
To overcome both issues, we only consider expressions written as the one of Eq.~\eqref{eq:MLTL},
where the dependence on causal propagators is explicitly pulled out. 

In a recent paper~\cite{Bobadilla:2021rmu}, we proposed an algorithm to generate causal 
representation of topologies constructed from $k+2$ cusps.
Due to the symmetry in the structure of  topologies when all possible connections, edges, 
between cusps are considered, we conjectured a close formula when 
all possible edges are considered,
\begin{align}
\mathcal{F}_{L+k} = & \sum_{\substack{i_{1}\ll i_{N_{i}}\\
j_{1}\ll j_{N_{j}}
}
}^{k+2}\Omega_{\vec{i}}^{\vec{j}}\ L_{i_{1}i_{2}\hdots i_{N_{i}}}^{+}L_{j_{1}j_{2}\hdots j_{N_{j}}}^{-}\,,
\label{eq:allcausal}
\end{align}
with, 
\begin{align}
\Omega_{\vec{i}}^{\vec{j}}&=
\begin{cases}
1 & \text{If }\vec{i}\cap\vec{j}=\emptyset\\
0 & \text{otherwise}
\end{cases}\,,
\end{align}
where $\vec{i}=\{i_1,i_2,\hdots,i_{N_i}\}$, $\vec{j}=\{j_1,j_2,\hdots,j_{N_j}\}$,
 $i_{1}\ll i_{N_{i}}$ is the lexicographic ordering $i_{1}<i_{2}<\cdots<i_{N_{i}}$,
 $N_{i}=\left[k/2\right]+1$ and $N_{j}=k-\left[k/2\right]$. 
The functions $L_{ijk\hdots}^{\pm}$ contain the causal information of the integrand,
\begin{align}
 & L_{i_{1}i_{2}\hdots i_{N}}^{\pm}=\frac{1}{\lambda_{i_{1}i_{2}\hdots i_{N}}^{\pm}}\sum_{\substack{j_{1}\ll j_{N-1}\\
\vec{j}\subset\vec{i}
}
}L_{j_{1}j_{2}\hdots j_{N-1}}^{\pm}\,, &  & L_{i_{1}}^{\pm}=\frac{1}{\lambda_{i_{1}}^{\pm}}\,.
\label{eq:allLs}
\end{align}
The close formula~\eqref{eq:allcausal}, together with the definition of $L^\pm$, have been implemented 
in \Lotty{} through the routine \texttt{AllCausal} by just typing the number of cusps of a given loop topology.

To illustrate the use of  \texttt{AllCausal}, let us consider the N$^2$MLT configuration,
described at the beginning of Sec.~\ref{sec:causal},  
{ \small
\begin{mmaCell}[moredefined={AllCausal}]{Input}
AllCausal[4]
\end{mmaCell}
}\vspace{-3ex}
{ \small
\begin{mmaCell}{Output}
  (L["-"][\{3\}] + L["-"][\{4\}]) L["+"][\{1, 2\}] 
+ (L["-"][\{2\}] + L["-"][\{4\}]) L["+"][\{1, 3\}] 
+ (L["-"][\{2\}] + L["-"][\{3\}]) L["+"][\{1, 4\}] 
+ (L["-"][\{1\}] + L["-"][\{4\}]) L["+"][\{2, 3\}] 
+ (L["-"][\{1\}] + L["-"][\{3\}]) L["+"][\{2, 4\}] 
+ (L["-"][\{1\}] + L["-"][\{2\}]) L["+"][\{3, 4\}] 
\end{mmaCell}
}
\noindent 
that completely agrees with the result of Ref.~\cite{Aguilera-Verdugo:2020kzc} when external momenta are attached to 
all cusps. In the notation of \Lotty, 
$L_k^- =$ \texttt{L["-"][\{k\}]} and $L_k^+ =$ \texttt{L["+"][\{k\}]}. 

Additionally, to have the expressions in terms of causal propagators, $\lambda_i^{\pm}$, 
one adds the flag \texttt{"ExpandToLambda"}, 
{ \small
\begin{mmaCell}[moredefined={AllCausal,tmp2,tmp3}]{Input}
tmp2 = AllCausal[4,"ExpandToLambda" -> True];
tmp3 = tmp2/. \(\lambda\)[_][a_] :> \(\lambda\)[a]
\end{mmaCell}
}\vspace{-3ex}
{ \small
\begin{mmaCell}{Output}
\mmaFrac{2 (\mmaFrac{1}{\(\lambda\)[\{3\}]}+\mmaFrac{1}{\(\lambda\)[\{2\}]}) (\mmaFrac{1}{\(\lambda\)[\{4\}]}+\mmaFrac{1}{\(\lambda\)[\{1\}]})}{\(\lambda\)[\{2,3\}]} + \mmaFrac{2 (\mmaFrac{1}{\(\lambda\)[\{2\}]}+\mmaFrac{1}{\(\lambda\)[\{1\}]}) (\mmaFrac{1}{\(\lambda\)[\{4\}]}+\mmaFrac{1}{\(\lambda\)[\{3\}]})}{\(\lambda\)[\{1,2\}]}
+\mmaFrac{2 (\mmaFrac{1}{\(\lambda\)[\{3\}]}+\mmaFrac{1}{\(\lambda\)[\{1\}]}) (\mmaFrac{1}{\(\lambda\)[\{4\}]}+\mmaFrac{1}{\(\lambda\)[\{2\}]})}{\(\lambda\)[\{1,3\}]}
\end{mmaCell}
}
\noindent 
Notice that while \texttt{tmp2} contains 24 ``causal'' terms,  \texttt{tmp3}
contains 12. This is because the energies of external momenta are set to zero,
corresponding to a vacuum like diagram,
whose numerical integration was performed in Ref.~\cite{Aguilera-Verdugo:2020kzc} in 
$d=2,3,4$ space-time dimensions. 

\smallskip 
In this section, we have discussed the parametric structure of the causal representation
of the most general topology in terms of $\lambda_{i}^{\pm}$
without making explicit the dependence on on-shell energies. 
Clearly, the latter depends on how the propagators for a given topology are defined
or, equivalently, how cusps are connected through edges.
Let us then elucidate the transition from the all-loop causal representation
to a particular one. 
To this end, we first consider the connection between the four cusps 
through the six edges (see Fig.~\ref{fig:3Ltopos}~(left)), 
{\small
\begin{mmaCell}[moredefined={q,SetSingLam}]{Input}
SetSingLam = \{
   q[\{1\}] -> \{1, 4, 6\},
   q[\{2\}] -> \{1, 2, 5\},
   q[\{3\}] -> \{3, 4, 5\},
   q[\{4\}] -> \{2, 3, 6\}
   \};
\end{mmaCell}
}\vspace{-1ex}
\noindent 
where $i$ in $q[\{i\}]$ corresponds to the $i$-th cusp in the topology
and is defined according to the lines that connect this cusp (red labels in Fig.~\ref{fig:3Ltopos}~(left)). 
Also, in order to pictorially see that the edges defined in \texttt{SetSingLam} are properly 
written, we implement the routine \texttt{PlotTop} that displays the topology under consideration,
{\small
\begin{mmaCell}[moredefined={PlotTop,SetSingLam}]{Input}
PlotTop[SetSingLam]
\end{mmaCell}
}\vspace{-1ex}
\begin{figure}[t]
\centering
\parbox{45mm}{\includegraphics[scale=0.45]{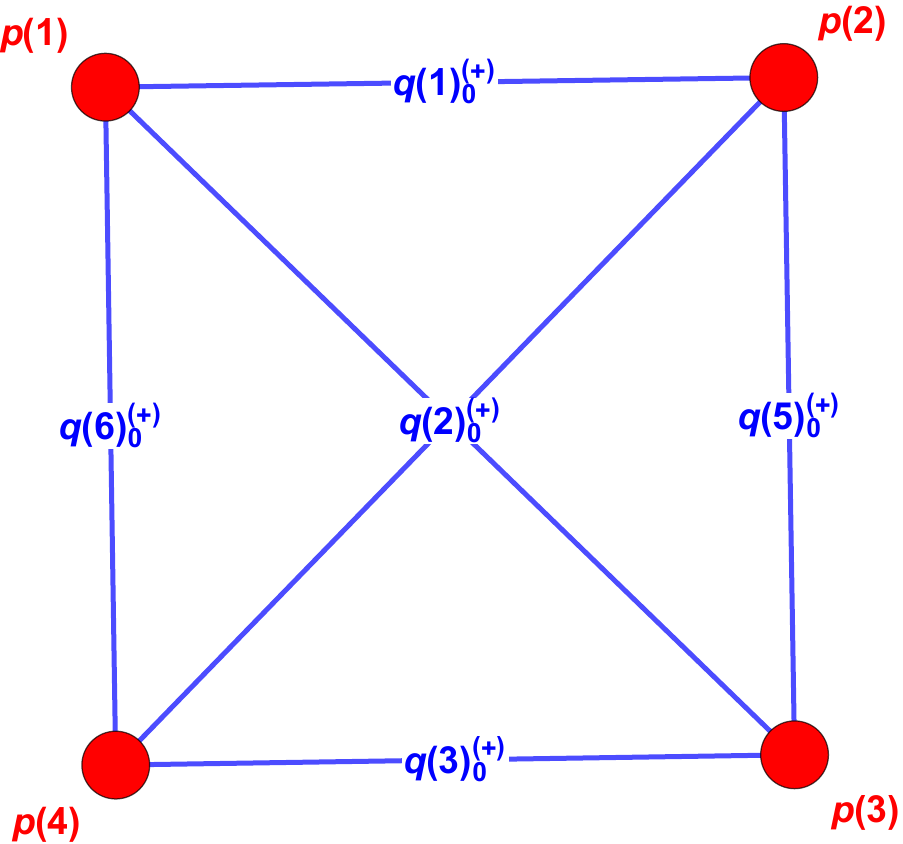}}
\parbox{50mm}{\includegraphics[scale=0.6]{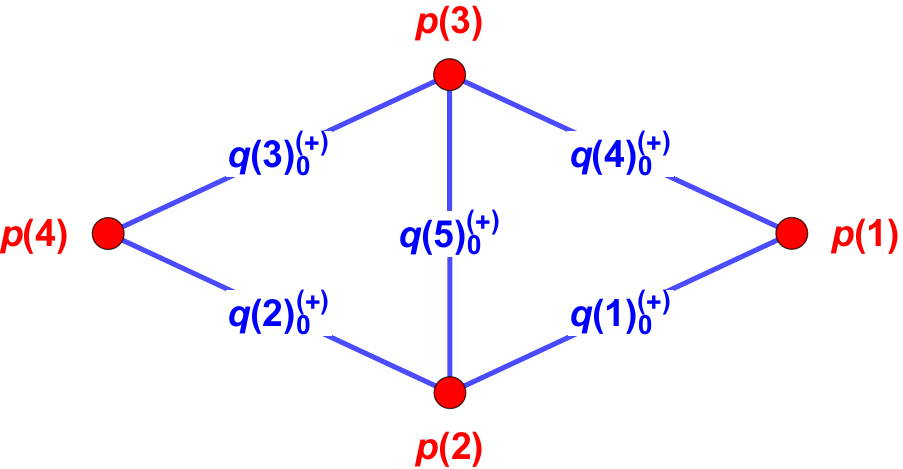}}
\qquad
\parbox{40mm}{\includegraphics[scale=0.45]{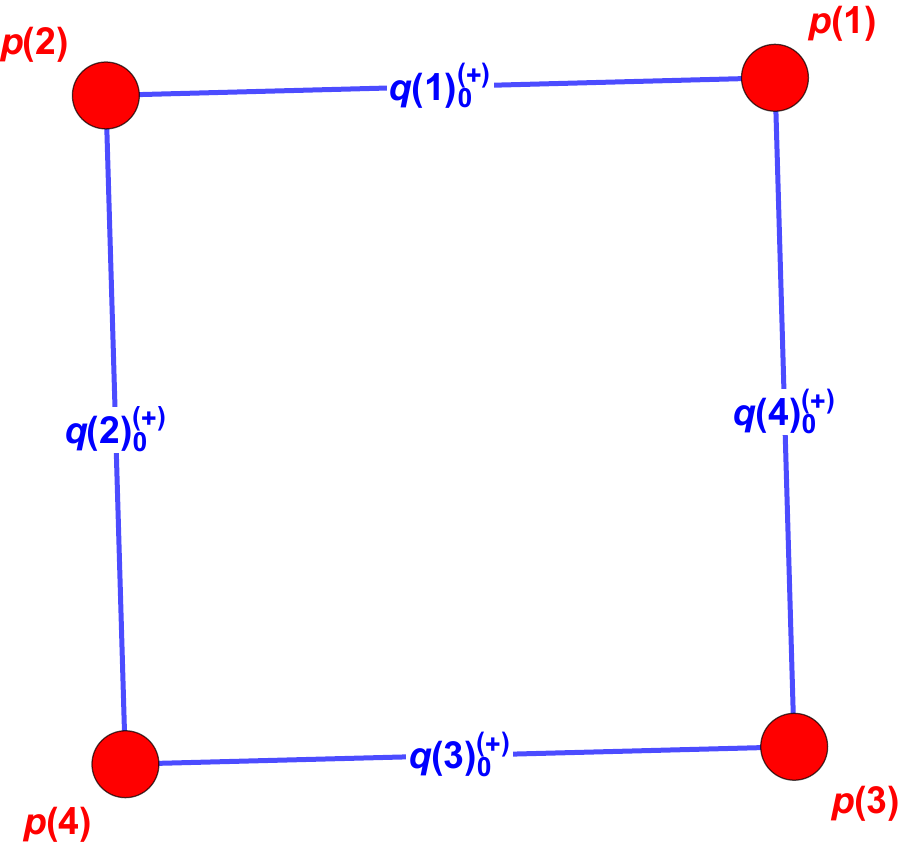}}
\caption{Loop topologies with four cusps generated with \texttt{PlotTop}.}
\label{fig:n2mlt2}
\end{figure}
\noindent 
producing a figure in a \Mathematica{} notebook, as the one shown in Fig.~\ref{fig:n2mlt2}~(left). 

The assignment of internal lines, according to the cusps, allows to generate the causal propagators
$\lambda_{i}^{\pm}$ in terms of $\qpw{i}$'s. 
This is internally carried out in \Lotty{} with the function \texttt{Lamb2qij}.
For instance, the causal representation of the 
three-loop 
N$^2$MLT configuration of Fig.~\ref{fig:n2mlt2}~(left) becomes, 
{\small
\begin{mmaCell}[moredefined={tmp3,AllCausal,momcon,toLambda,tmp4,SetSingLam,Lamb2qij}]{Input}
tmp3 = AllCausal[4, "ExpandToLambda" -> True];
momcon = \{p[4] -> - p[1] - p[2] - p[3]\};
toLambda = Lamb2qij[tmp3, SetSingLam];
tmp4 = tmp3 //. toLambda /. momcon;
\end{mmaCell}
}\vspace{-1ex}
\noindent
In \texttt{tmp3}, we generate the most general causal representation for a topology 
built from four edges. Then, in \texttt{momcon}, we define the momentum conservation
of the four external particles. The routine \texttt{Lamb2qij}, as explained above,
performs the replacement from $\lambda_{i}$'s to $\qpw{(i_1,i_2,\hdots)}$. 
Finally, in the last line, we apply all the conditions, finding the very same structure 
of the N$^2$MLT configuration of Ref.~\cite{Aguilera-Verdugo:2020kzc}. 

On top of the representation found for the three-loop topology,
as discussed in Ref.~\cite{Bobadilla:2021rmu}, one can profit from the causal
representation of the most general topology, in this case N$^2$MLT, 
and obtaining without any evaluation of the residue the causal representation
of topologies with lower number of edges, but same number of cusps. 
In effect, we can also generate topologies with five (Fig.~\ref{fig:n2mlt2}~(center)) 
and four (Fig.~\ref{fig:n2mlt2}~(right)) edges,
by removing $\qpw{6}$ and $\qpw{5},\qpw{6}$, respectively,
{\small
\begin{mmaCell}[moredefined={q,SetSingLam1,SetSingLam2,SetSingLam3}]{Input}
SetSingLam2 = \{
   q[\{1\}] -> \{1, 4\},
   q[\{2\}] -> \{1, 2, 5\},
   q[\{3\}] -> \{3, 4, 5\},
   q[\{4\}] -> \{2, 3\}
   \};
SetSingLam3 = \{
   q[\{1\}] -> \{1, 4\},
   q[\{2\}] -> \{1, 2\},
   q[\{3\}] -> \{3, 4\},
   q[\{4\}] -> \{2, 3\}
   \};   
\end{mmaCell}
}\vspace{-1ex}\noindent
Then, to obtain the causal representation of a two-loop kite diagram 
(Fig.~\ref{fig:n2mlt2}~(middle)) with two external momenta, 
one focuses only on the connection of edges and momentum conservation,
{\small
\begin{mmaCell}[moredefined={tmp3,AllCausal,momcon,toLambda,tmp4,SetSingLam2,Lamb2qij}]{Input}
tmp3 = AllCausal[4, "ExpandToLambda" -> True];
momcon = \{p[2] -> - p[1], Subscript[p[3 | 4], _] :> 0 \};
toLambda = Lamb2qij[tmp3, SetSingLam2];
tmp4 = tmp3 //. toLambda /. momcon ;
\end{mmaCell}
}\vspace{-1ex}
\noindent
where the momenta $p_3$ and $p_4$, associated to the cusps $3$ and $4$, respectively,
are set to zero to satisfy momentum conservation. 

Similarly, for the one-loop box, 
{\small
\begin{mmaCell}[moredefined={tmp3,AllCausal,momcon,toLambda,tmp4,SetSingLam3,Lamb2qij}]{Input}
tmp3 = AllCausal[4, "ExpandToLambda" -> True];
momcon = \{p[4] -> - p[1] - p[2] - p[3]\};
toLambda = Lamb2qij[tmp3, SetSingLam3];
tmp4 = tmp3 //. toLambda /. momcon;
\end{mmaCell}
}\vspace{-1ex}
Let us remark that, as explained in Ref.~\cite{Bobadilla:2021rmu}, cancellations 
due to the absence of $\qpw{5}$ and/or $\qpw{6}$
appear. In particular, in the former cases the number of ``causal terms'' decreases from 24 to 20.

\subsection{\texttt{RefineCausal}}

It is clear that when generating the most general causal representation of a loop topology, 
given by a fixed number of cusps, 
several redundant terms will appear. 
Hence, to compensate their appearance, we work out the expressions of $L^{\pm}$ 
before expanding in $\lambda^{\pm}$'s. 
This is straightforwardly carried out by studying the connections between cusps,
provided e.g. by \texttt{SetSingLam2} and \texttt{SetSingLam3},
through the routine  \texttt{RefineCausal}. 

To illustrate the use of \texttt{RefineCausal}, 
let us keep studying the topologies with four cusps of Fig~\ref{fig:n2mlt2},
where one- and two-loop topologies are particular cases of the 
N$^2$MLT configuration. 
Then, starting from the general causal representation, 
we explicitly add the connections between cusps and edges for 
the two-loop topology as follows, 
{\small
\begin{mmaCell}[moredefined={AllCausal,tmp5,RefineCausal,SetSingLam2}]{Input}
tmp5 = AllCausal[4] // RefineCausal[#, SetSingLam2] &
\end{mmaCell}
}\vspace{-3ex}
{ \small
\begin{mmaCell}{Output}
  (L["-"][\{3\}] + L["-"][\{4\}]) L["+"][\{1, 2\}] 
+ (L["-"][\{2\}] + L["-"][\{4\}]) L["+"][\{1, 3\}] 
+ (L["-"][\{2\}] + L["-"][\{3\}]) L["+"][\{1\}] L["+"][\{4\}] 
+ (L["-"][\{1\}] + L["-"][\{4\}]) L["+"][\{2, 3\}]
+ (L["-"][\{1\}] + L["-"][\{3\}]) L["+"][\{2, 4\}] 
+ (L["-"][\{1\}] + L["-"][\{2\}]) L["+"][\{3, 4\}] 
\end{mmaCell}
}
\noindent
This result is expected 
since the cusps 1 and 4 do not share any edge, as noticed in the 
definition of \texttt{SetSingLam2}, and allows for reducing from 24 to 22 causal terms. 
Besides, in this configuration, due mainly to momentum conservation 
of the external momenta, we have a further relation, 
\begin{align}
(L^{-}_{1}+L^{-}_{4})L^{+}_{23}=(L^{+}_{2}+L^{+}_{3})L^{-}_{1}L^{-}_{4}\,,
\end{align}
by taking into account the explicit dependence on $\lambda^{\pm}$ in Eq.~\eqref{eq:allLs}
and $\lambda_{14}^{\pm} = \lambda_{23}^{\mp}$.
Thus, recovering the 20 causal terms of Ref.~\cite{Bobadilla:2021rmu}.

Similar for the one-loop box, 
{\small
\begin{mmaCell}[moredefined={AllCausal,tmp5,RefineCausal,SetSingLam3}]{Input}
tmp5 = AllCausal[4] // RefineCausal[#, SetSingLam3] &
\end{mmaCell}
}\vspace{-3ex}
{ \small
\begin{mmaCell}{Output}
  (L["-"][\{3\}] + L["-"][\{4\}]) L["+"][\{1, 2\}] 
+ (L["-"][\{2\}] + L["-"][\{4\}]) L["+"][\{1, 3\}] 
+ (L["-"][\{2\}] + L["-"][\{3\}]) L["+"][\{1\}] L["+"][\{4\}] 
+ (L["-"][\{1\}] + L["-"][\{4\}]) L["+"][\{2\}] L["+"][\{3\}]
+ (L["-"][\{1\}] + L["-"][\{3\}]) L["+"][\{2, 4\}] 
+ (L["-"][\{1\}] + L["-"][\{2\}]) L["+"][\{3, 4\}] 
\end{mmaCell}
}\noindent
we notice $L_{23}\to L_{2}^{\pm}L_{3}^{\pm}$ because cusps 2 and 3 are not 
connected by any edge. 

Since the main idea of the causal representation is the numerical evaluation, 
one can always numerically compare the values of the sum of all residues,
yet contaminated by non-causal propagators, 
with the one obtained from the all causal representation proposed in Ref.~\cite{Bobadilla:2021rmu}. 
Clearly, the former will display several numerical instabilities w.r.t. the latter.
Then to avoid this issue, we evaluate both integrands in the non-physical region,
$\text{Im}(p_i)\ne0$.

In order to compare both approaches, let us focus on the one-loop box of Fig.~\ref{fig:n2mlt2} (right), 
{\small
\begin{mmaCell}[moredefined={propagators,assumptions}]{Input}
propagators = \{l1, l1+p[1], l1+p[1]+p[2], l1+p[1]+p[2]+p[3]\};
assumptions = (Im[Subscript[p[#], 0]] == 0) & /@ Range[3];
\end{mmaCell}
}\vspace{-1ex}
\noindent 
and generate the dual representation of this integrand, 
{\small
\begin{mmaCell}[moredefined={tmp2,tmp,propagators,assumptions,loopmom,GetDual,GetCausalProps}]{Input}
tmp  = GetDual[1, propagators, loopmom,"Assumptions" -> assumptions];
tmp2 = tmp*GetCausalProps[tmp, propagators, "GetPref" -> True] // Total; 
 \end{mmaCell}
}%\vspace{-1ex}
\noindent 
as well as the causal one, 
{\small
\begin{mmaCell}[moredefined={tmp3,AllCausal,momcon,toLambda,tmp4,SetSingLam2,Lamb2qij,RefineCausal}]{Input}
tmp3 = AllCausal[4]// RefineCausal[#, SetSingLam2] &;
momcon = \{p[4] -> -p[1] - p[2] - p[3]\};
toLambda = Lamb2qij[tmp3, SetSingLam2];
tmp4 = tmp3 //. toLambda /. momcon ;
\end{mmaCell}
}\vspace{-1ex}
\noindent 
Thus, the numerical evaluation of both integrands can be performed and compared as follows, 
{ \small
\begin{mmaCell}[moredefined={tmp2,tmp4,repl1,momcon,RandomInteger,propagators}]{Input}
repl1 = (Subsuperscript[q[#], 0, "(+)"] -> 
		\mmaFrac{RandomInteger[\{10^3, 10^4\}]}{RandomInteger[\{10^4, 10^5]\}]}) & /@Range[Length@propagators];
		
repl1 = Join[repl1, ( Subscript[p[#], 0] -> 
		\mmaFrac{RandomInteger[\{10^3, 10^4\}]}{RandomInteger[\{10^4, 10^5]\}]} I) & /@Range[3]];
		
tmp2 + tmp4 /. momcon /. repl1
\end{mmaCell}
}\vspace{-3ex}
{ \small
\begin{mmaCell}{Output}
0
\end{mmaCell}
}

\noindent
Notice that, in order to find a numerical agreement between causal 
and non-causal (sum of residues) representation,
we profit from the evaluation of the integrand over random rational points. 
The same approach is carried out for all the examples collected in \texttt{usage.nb}. 

\subsection{Six-cusp topologies}

\begin{figure}[t]
\centering
\parbox{40mm}{\includegraphics[scale=0.3]{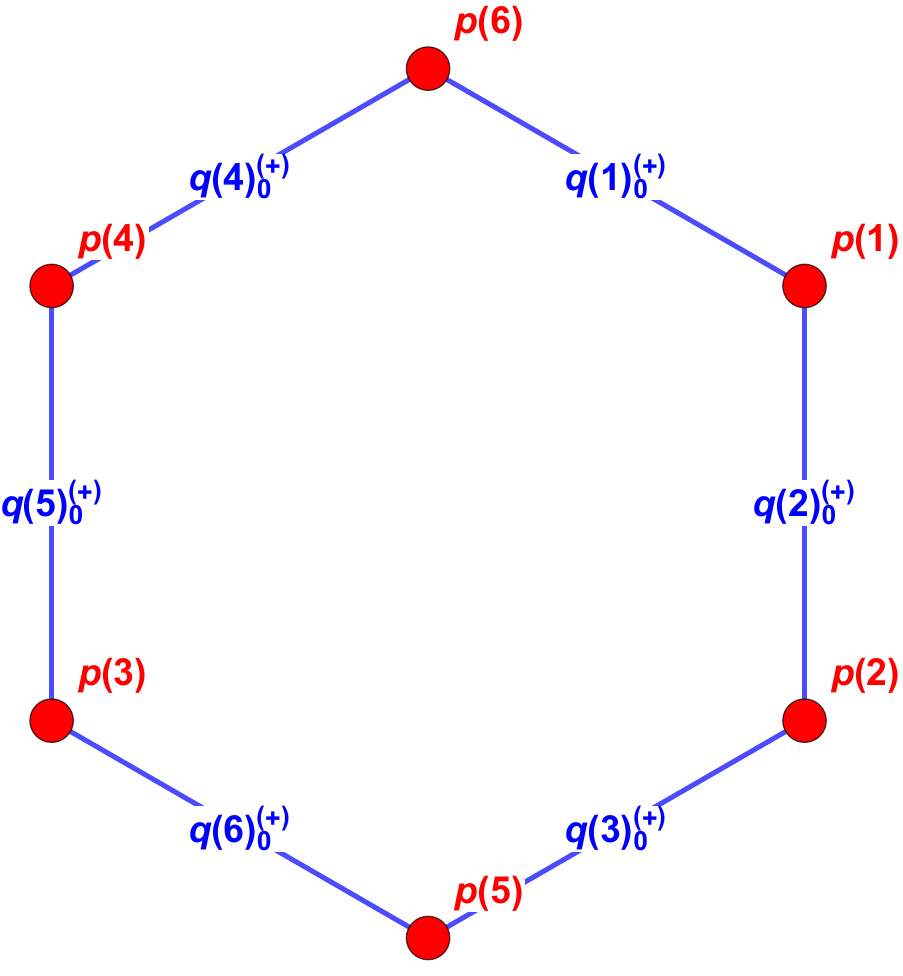}}
\parbox{45mm}{\includegraphics[scale=0.5]{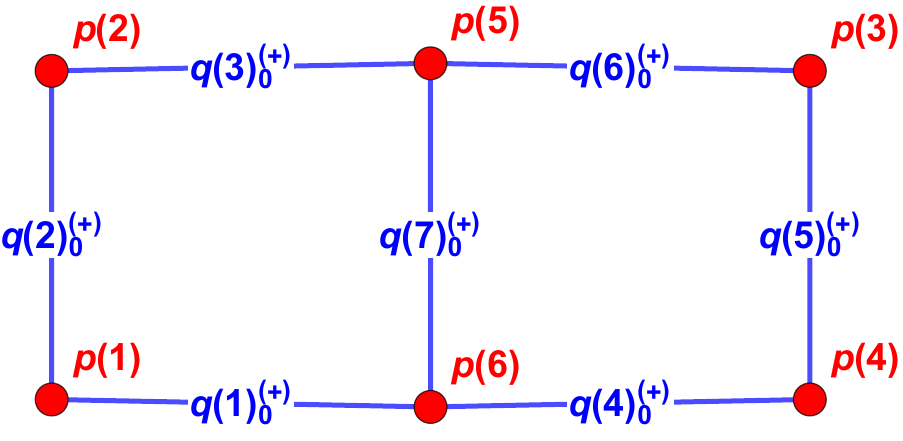}}
\qquad
\parbox{50mm}{\includegraphics[scale=0.6]{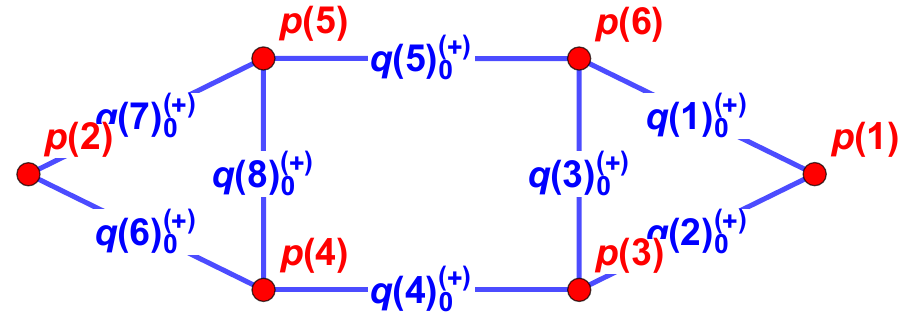}}
\caption{Loop topologies with six cusps generated with \texttt{PlotTop}.}
\label{fig:n4mlt}
\end{figure}
In the previous section, we discussed that the routine \texttt{AllCausal}
allows to generate the most
general causal representation of a loop topology, given by the number of cusps
and their connections through edges.
Remarkably, the use of LTD is not needed to generate those expressions,
but it is important to numerically compare both approaches.  
Hence, to elucidate this advantage, in this section we present the causal 
representation of loop topologies made of six cusps.

The general causal representation of a topology with six cusps
generates topologies up-to fifteen loops. In Fig.~\ref{fig:n4mlt}, 
we display a few topologies (up-to three loops) that can be obtained
from the most general one. 
In the following, we elaborate on the two-loop double box.
The other topologies at one and three loops are analysed 
in the notebook \texttt{usage.nb},
following the same approach as this section. 

Let us then start with the list of \texttt{propagators}
of the two-loop double box of Fig.~\ref{fig:2Ltopos}~(right)
{\small
\begin{mmaCell}[moredefined={propagators}]{Input}
propagators = \{l1, l1+p[1], l1+p[1]+p[2], l2, l2+p[4], l2+p[3]+p[4], -l1-l2\};
\end{mmaCell}
}\vspace{-1ex}
\noindent 
where, by means of cusps and edges, this topology can be understood as,
{\small
\begin{mmaCell}[moredefined={q,SetSingLam}]{Input}
SetSingLam = \{
   q[\{1\}] -> \{1, 2\},
   q[\{2\}] -> \{2, 3\},
   q[\{3\}] -> \{6, 5\},
   q[\{4\}] -> \{5, 4\},
   q[\{5\}] -> \{3, 7, 6\},
   q[\{6\}] -> \{1, 4, 7\}
   \};
\end{mmaCell}
}\vspace{-1ex}
\noindent 
The edges in \texttt{SetSingLam} are defined according to Fig.~\ref{fig:2Ltopos}~(right)
and generates, through \texttt{PlotTop[SetSingLam]}, Fig.~\ref{fig:n4mlt}~(middle). 
Thus, the causal representation of the scalar two-loop double box
is found to be, 
{\small
\begin{mmaCell}[moredefined={tmp3,AllCausal,momcon,toLambda,tmp4,SetSingLam,Lamb2qij}]{Input}
tmp3 = AllCausal[6, "ExpandToLambda" -> True];
momcon = \{p[4] -> -p[1] - p[2] - p[3], Subscript[p[5 | 6], _] :> 0\};
toLambda = Lamb2qij[tmp3, SetSingLam];
tmp4 = tmp3 //. toLambda /. momcon ;
\end{mmaCell}
}\vspace{-1ex}
\noindent
Similar to the one- and two-loop topologies considered in the previous section, 
in \texttt{tmp3}, we generate the most general causal representation for a topology 
built from six cusps. Then, in \texttt{momcon}, we define the momentum conservation
of the four external particles, setting to zero 
 those that are not present in this configuration, $p_5$ and $p_6$. 
The routine \texttt{Lamb2qij}, as explained above,
performs the replacement from $\lambda_{i}$'s to $\qpw{(i_1,i_2,\hdots)}$. 
Finally, in the last line, we apply all the conditions.
 
Since the full value of \texttt{tmp4} is not illuminating for the current discussion, 
we just verify the variable the latter depends on,
{\small
\begin{mmaCell}[moredefined={tmp4}]{Input}
tmp4 // Variables // Sort
\end{mmaCell}
}\vspace{-3ex}
{\small
\begin{mmaCell}[]{Output}
\{ \mmaSub{p[1]}{0}, \mmaSub{p[2]}{0}, \mmaSub{p[3]}{0}, \mmaSubSup{q[1]}{0}{(+)}, \mmaSubSup{q[2]}{0}{(+)}, \mmaSubSup{q[3]}{0}{(+)}, \mmaSubSup{q[4]}{0}{(+)}, \mmaSubSup{q[5]}{0}{(+)}, \mmaSubSup{q[6]}{0}{(+)}, \mmaSubSup{q[7]}{0}{(+)} \}
\end{mmaCell}
}\vspace{-1ex}\noindent
which is expected from the LTD formalism. 

In order to check that the stored expression in \texttt{tmp4} corresponds 
to a better rewriting of the dual integrands in terms of causal propagators, 
we compare this result with 
the one obtained directly from the nested application
of the residue~\eqref{eq:nestres}.
%Although the latter is contaminated by non-causal propagators
%and displays numerical instabilities.
To this end, we generate the residues in the standard way, 
as done in Sec.~\ref{sec:db_res}, 
{\small
\begin{mmaCell}[moredefined={tmp2,tmp,propagators,assumptions,loopmom,GetDual,GetCausalProps}]{Input}
tmp   = GetDual[1, propagators, loopmom,"Assumptions" -> assumptions];
tmp2 = tmp*GetCausalProps[tmp, propagators, "GetPref" -> True] // Total; 
 \end{mmaCell}
}\vspace{-1ex}
\noindent 
and evaluate $\qpw{i}$'s and $p_i$'s in random rational points,
{\small
\begin{mmaCell}[moredefined={tmp2,tmp4,momcon}]{Input}
tmp2/.momcon/._Subscript->I/.Subsuperscript[q[i_],__]:>i//AbsoluteTiming
tmp4/._Subscript->I/.Subsuperscript[q[i_],__]:>i//AbsoluteTiming
%%/% // Last
 \end{mmaCell}
}\vspace{-3ex}
{\small
\begin{mmaCell}[]{Output}
\{0.006438, -(1108252453/288937610000)\}
\end{mmaCell}
}\vspace{-3ex}
{\small
\begin{mmaCell}[]{Output}
\{0.00648, 1108252453/288937610000\}
\end{mmaCell}
}\vspace{-3ex}
{\small
\begin{mmaCell}[]{Output}
-1
\end{mmaCell}
}\vspace{-1ex}
\noindent
whose difference in both evaluations is an overall sign, which can be accounted for in 
the general expression of Eq.~\eqref{eq:allcausal}.

 \section{Numerical integration}
\label{sec:numeval}

In the previous sections, we discussed and provided automated tools to generate integrands
in the causal representation. As mentioned throughout this paper and in former works, 
the aim of  doing so is to be able to perform smooth and efficient numerical evaluations of the latter. 
Then, with this approach in mind, we provide non-trivial examples of the evaluation
of three-point finite integrals at two loops. 
Differently from previous studies, we consider multi-loop Feynman integrands with presence
of external momenta, yet getting agreement with publicly available softwares based on 
sector decomposition. 

Since the main feature of \Lotty{} is performing as many symbolic operations as possible
to render in a very simple integrand, 
we implement, for numerical purposes, the change of variables discussed in Ref.~\cite{Aguilera-Verdugo:2020kzc},
which we recall presently. 

\subsection{Change of variables}
\label{sec:changevars}

In order to perform a numerical integration of an $L$-loop Feynman integrand, 
one has to deal, in general, with $(d-1) L$ integrations,
which can be straightforwardly embedded in $(d-1) L$-dimensional hypersphere
or in the product of $L$  $(d-1)$-dimensional spheres.\footnote{For
an extensive discussion on the choice of the integration variables, 
we refer the reader to Ref.~\cite{Aguilera-Verdugo:2020kzc}.} 
The former and latter change of variables are implemented in \Lotty{}
by activating the flag \texttt{"Domain"}
with the options \texttt{"HyperSphere"} and \texttt{"Sphere"}, respectively,
in the routines \texttt{LoopToSC}, \texttt{meassure} and \texttt{intvars}. 
Let us elucidate their use with an example of an integrand at two loops
in four space-time dimensions, 
{\small
\begin{mmaCell}[moredefined={LoopMom,dim}]{Input}
dim = 4;		  (* set the dimension *)
LoopMom = \{l1, l2\};	(* list of loop momenta *)
\end{mmaCell}
}\vspace{-3ex}
{\small
\begin{mmaCell}[moredefined={LoopMom,dim,LoopToSC,meassure,intvars}]{Input}
LoopToSC[LoopMom, dim, "Domain" -> "HyperSphere"]
meassure[LoopMom, dim, "Domain" -> "HyperSphere"]
intvars[LoopMom, dim,  "Domain" -> "HyperSphere", "MaxValueR"->max]
\end{mmaCell}
}\vspace{-3ex}
{\small
\begin{mmaCell}{Output}
\{l1 -> \{r1 Cos[\(\theta\)1], r1 Cos[\(\theta\)2]*Sin[\(\theta\)1], r1 Cos[\(\theta\)3] Sin[\(\theta\)1] Sin[\(\theta\)2]\}, 
 l2 -> \{r1 Cos[\(\theta\)4] Sin[\(\theta\)1] Sin[\(\theta\)2] Sin[\(\theta\)3],
	r1 Cos[\(\theta\)5] Sin[\(\theta\)1] Sin[\(\theta\)2] Sin[\(\theta\)3] Sin[\(\theta\)4],
	r1 Sin[\(\theta\)1] Sin[\(\theta\)2] Sin[\(\theta\)3] Sin[\(\theta\)4] Sin[\(\theta\)5] \} \}
\end{mmaCell}
}\vspace{-3ex}
{\small
\begin{mmaCell}{Output}
\mmaSup{r1}{5} \mmaSup{Sin[\(\theta\)1]}{4} \mmaSup{Sin[\(\theta\)2]}{3} \mmaSup{Sin[\(\theta\)3]}{2} Sin[\(\theta\)4]
\end{mmaCell}
}\vspace{-3ex}
{\small
\begin{mmaCell}{Output}
\{\{r1,0,max\}, \{\(\theta\)1,0,\(\pi\)\}, \{\(\theta\)2,0,\(\pi\)\}, \{\(\theta\)3,0,\(\pi\)\}, \{\(\theta\)4,0,\(\pi\)\}, \{\(\theta\)5,0,2\(\pi\)\}\}
\end{mmaCell}
}
\noindent 
Similarly, the change of variables for the products of $L$ $(d-1)$-dimensional spheres, 
{\small
\begin{mmaCell}[moredefined={LoopMom,dim,LoopToSC,meassure,intvars}]{Input}
LoopToSC[LoopMom, dim, "Domain" -> "Sphere"]
meassure[LoopMom, dim, "Domain" -> "Sphere"]
intvars[LoopMom, dim,  "Domain" -> "Sphere", "MaxValueR"->max]
\end{mmaCell}
}\vspace{-3ex}
{\small
\begin{mmaCell}{Output}
\{l1 -> \{r1 Cos[\(\theta\)11], r1 Cos[\(\theta\)12] Sin[\(\theta\)11],\
r1 Sin[\(\theta\)11] Sin[\(\theta\)12] \}, 
 l2 -> \{r2 Cos[\(\theta\)21], r2 Cos[\(\theta\)22] Sin[\(\theta\)21],\ 
r2 Sin[\(\theta\)21] Sin[\(\theta\)22] \} \}
\end{mmaCell}
}\vspace{-3ex}
{\small
\begin{mmaCell}{Output}
\mmaSup{r1}{2} \mmaSup{r2}{2} Sin[\(\theta\)11] Sin[\(\theta\)21]
\end{mmaCell}
}\vspace{-3ex}
{\small
\begin{mmaCell}{Output}
\{\{r1,0,max\}, \{r2,0,max\}, \{\(\theta\)11,0,\(\pi\)\}, \{\(\theta\)12,0,\
2\(\pi\)\}, \{\(\theta\)21,0,\(\pi\)\}, \{\(\theta\)22,0,2\(\pi\)\}\}   
\end{mmaCell}
}\vspace{-1ex}
\noindent
Let us comment on the routines  we introduced in the previous example.
\texttt{LoopToSC} provides the replacement rules to spherical 
coordinates for the spatial components of the loop momenta
as  $(d-1)$ arrays for each loop.
\texttt{meassure} and \texttt{intvars} write, 
according to the chosen \texttt{"Domain"}, 
the integration measure and integration variables, respectively.
Notice that, due to the LTD formalism, the simplicity in this change of variables stems 
of the transition from Minkowski to Euclidean space. 

The default domain of the above routines is \texttt{"Sphere"}, 
which, as mentioned before, corresponds to embed all integrations in a product of $L$ $(d-1)$-dimensional
spheres. In this case, the notation for angles~$\theta_{ij}$ and radius~$r_i$ 
exactly corresponds to the $i$-th loop momentum in the list \texttt{LoopMom}. 
In the integration limits of each radius, one can choose the upper bound, 
by default we set \texttt{"MaxValueR"->$10^5$}. 
Nevertheless, to avoid considering the integration domain $[0,\infty)$,
one can map $[0,\infty)$ to the segment $(0,1]$, 
with the change of variables, 
\begin{align}
r_i = \frac{1-x_i}{x_i}\,,
\label{eq:cpmpact}
\end{align}
making thus the numerical evaluation of the integrand simpler. 

\begin{figure}[t]
\centering
\parbox{30mm}{\includegraphics[scale=1.2]{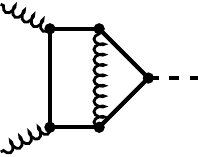}}
\parbox{30mm}{\includegraphics[scale=1.2]{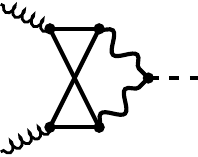}}
\caption{Representative two-loop planar (left) and non-planar (right) triangle diagrams.
Solid and wavy lines correspond to massive propagators with different masses;
wavy lines correspond to massless particles. 
}
\label{fig:triangles}
\end{figure}

Since the integrands are written in terms of the on-shell energies $\qpw{i}$,
we provide an additional routine to perform the scalar products in the  
Euclidean space, \texttt{sp[a,b]}. 
For instance, the scalar product of $\boldsymbol{\ell}_{1}\cdot\boldsymbol{\ell}_{2}$
in $(4-1)$ dimensions, 
{\small
\begin{mmaCell}[moredefined={dim,sp,LoopToSC}]{Input}
dim = 4;
sp[l1, l2] /. LoopToSC[\{l1,l2\}, dim] // FullSimplify
\end{mmaCell} 
}\vspace{-3ex}
{\small
\begin{mmaCell}{Output}
r1 r2 (Cos[\(\theta\)11] Cos[\(\theta\)21] + Cos[\(\theta\)12 - \(\theta\)22] Sin[\(\theta\)11] Sin[\(\theta\)21])
\end{mmaCell}
}\vspace{-1ex}\noindent
where, to evaluate these scalar products, one needs to add two lists (with the same length).
In the former example, three-dimensional arrays for $\boldsymbol{\ell}_{1}$ and $\boldsymbol{\ell}_{2}$
are given by the parametrisation, in spherical coordinates, provided by \texttt{LoopToSC[{l1, l2}, dim]},
as mentioned above. 

Hence, with the routine \texttt{sp} ready to perform $(d-1)$-dimensional scalar products 
in Euclidean space, we provide, for illustrative reasons, 
the on-shell energies of the two-loop planar triangle of 
Fig.~\ref{fig:triangles}~(left),
{\small
\begin{mmaCell}[moredefined={LoopMom,spatial,dim,propagators,energies}]{Input}
dim = 4;
energies = \{Subscript[p[1],0]->Ecm/2, Subscript[p[2],0]->Ecm/2\};
spatial  = \{p[1]->Ecm/2 \{1,0,0\}, p[2]->Ecm/2 \{-1,0,0\}\};
LoopMom = \{l1,l2\};
propagators = \{l1,l1+p[1],l1+p[1]+p[2],l2,l2-p[1]-p[2],l1+l2\};
\end{mmaCell}
}\noindent
Notice that the user has to include the four-dimensional arrays for the external momenta. 
In this particular example, the external momenta are chosen to be in the 
center-of-mass frame (COM). Thus, 
\texttt{Ecm} corresponds to the COM energy,
yielding to the 
on-shell energies in spherical coordinates,
{\small
\begin{mmaCell}[moredefined={LoopMom,spatial,dim,propagators,myqi0,value,myrepl,LoopToSC,sp}]{Input}
myqi0 = Subsuperscript[q[#],0,"(+)"]&/@Range[Length@propagators];
value = myqi0/.Subsuperscript[q[ii_],__]:>Sqrt[sp@propagators[[ii]]+m[ii]^2];
value = value/.LoopToSC[LoopMom,dim]/.spatial;
myrepl = Thread[myqi0->value]//FullSimplify
\end{mmaCell}
}\vspace{-3ex}
{\small
\begin{mmaCell}{Output}
\{\mmaSubSup{q[1]}{0}{(+)} ->  \mmaSqrt{\mmaSup{r1}{2} + \mmaSup{m[1]}{2}}, 
 \mmaSubSup{q[2]}{0}{(+)} -> \mmaFrac{1}{2}\mmaSqrt{\mmaSup{Ecm}{2} + 4 Ecm r1 Cos[\(\theta\)11] + 4 (\mmaSup{r1}{2} + \mmaSup{m[2]}{2})}, 
 \mmaSubSup{q[3]}{0}{(+)} ->  \mmaSqrt{\mmaSup{r1}{2} + \mmaSup{m[3]}{2}}, 
 \mmaSubSup{q[4]}{0}{(+)} ->  \mmaSqrt{\mmaSup{r2}{2} + \mmaSup{m[4]}{2}}, 
 \mmaSubSup{q[5]}{0}{(+)} ->  \mmaSqrt{\mmaSup{r2}{2} + \mmaSup{m[5]}{2}}, 
 \mmaSubSup{q[6]}{0}{(+)} ->  \mmaSqrt{\mmaSup{r1}{2}+\mmaSup{r2}{2}+\mmaSup{m[6]}{2}+2r1r2 (Cos[\(\theta\)11]Cos[\(\theta\)21] + Cos[\(\theta\)12-\(\theta\)22]Sin[\(\theta\)11]Sin[\(\theta\)21])} \}
\end{mmaCell}
}\noindent
where in the temporary variable \texttt{value} we make the replacement assuming
that the mass of the $i$-th propagator is $m_i$. 

In the derivation of the on-shell energies, let us note that one
can harness from the choice of the reference frame.
For instance, in the evaluation of the two-loop planar triangle of Fig.~\ref{fig:triangles}~(left),
one has that the only massless propagator is $\ell_{1}+\ell_{2}$
while the other ones are massive with the same mass. 
Then, working in the COM frame, a simplification at the level of on-shell energies 
is manifest, $\qpw{3}=\qpw{1}$ and $\qpw{5} =\qpw{4}$. 
This choice of the reference frame reduces the number of variables to evaluate from 
6 to 4. 
In the following section, we discuss the stability of the causal representation of this integrand. 

\subsection{Numerical stability}

Then, with $\qpw{i}$ expressed in spherical coordinates, 
we can study  the numerical stability 
of integrands in the causal representation. 
To do so, we consider 
the two-loop planar and non-planar scalar triangles depicted in Fig.~\ref{fig:triangles},
whose complete study of the renormalised amplitudes have been carried out 
in Refs.~\cite{Driencourt-Mangin:2019aix,Driencourt-Mangin:2019yhu}
for the Higgs production in QED and in QCD/EW, respectively.  

In the following examples, we work in the region where no threshold singularities 
appear in the evaluation of the integrand. 
Nevertheless, the extension to integrands containing thresholds can yet be carried out
with the aid of the routine \texttt{GetCausalProps} presented in Sec.~\ref{sec:causal},
since the latter explicitly displays the causal propagators and thus 
the locus of all thresholds. 
Then, we specify the mass of  \texttt{propagators} according to  Fig.~\ref{fig:triangles},
where, for the planar triangle, the propagator with a curly line is considered as massless
whereas the other ones are massive. 
For the non-planar triangle, we choose all masses to be the same. 
We recall that choosing different values of masses for the internal lines
(solid and wavy) do not generate any limitation to our approach, 
since we perform a purely numerical evaluation, as observed in 
Ref.~\cite{Driencourt-Mangin:2019yhu}.  

\begin{figure}[t!]
\includegraphics[scale=0.85]{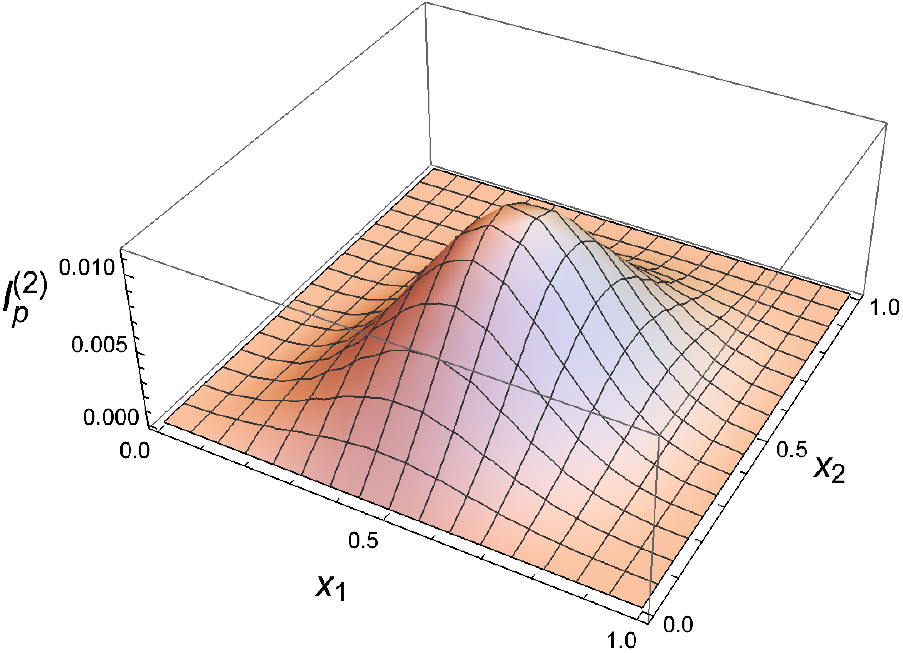}\quad
\includegraphics[scale=0.85]{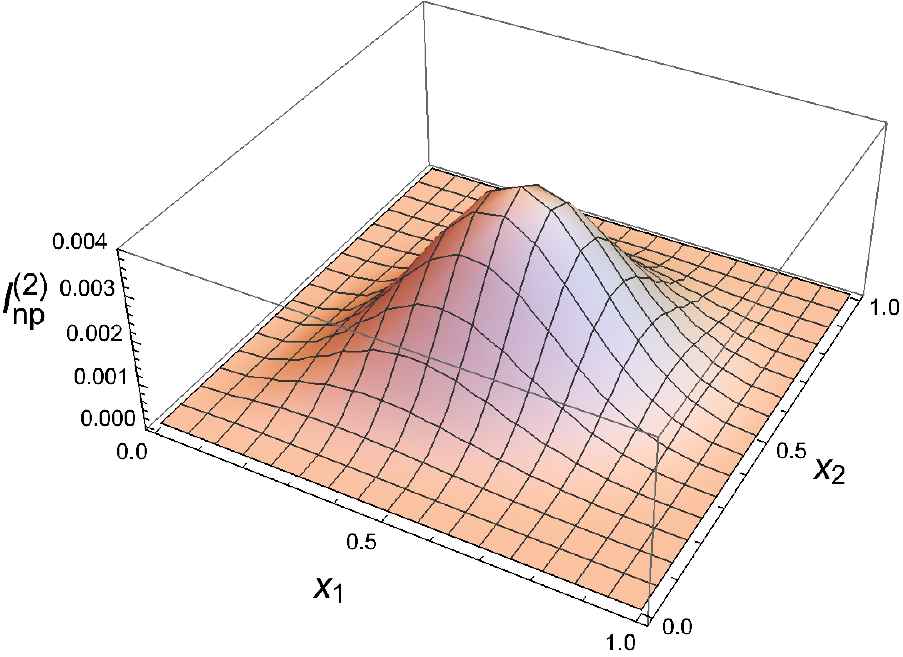}
\caption{Numerical stability of the integrands of
the two-loop planar (left) and non-planar (right) scalar triangles of Fig.~\ref{fig:triangles}
as a function of $x_1$ and $x_2$ with fixed values for the angles $\theta_{ij}$. 
}
\label{fig:3d_tris}
\end{figure}
\begin{figure}[t!]
(a)\quad\parbox{75mm}{\includegraphics[scale=0.7]{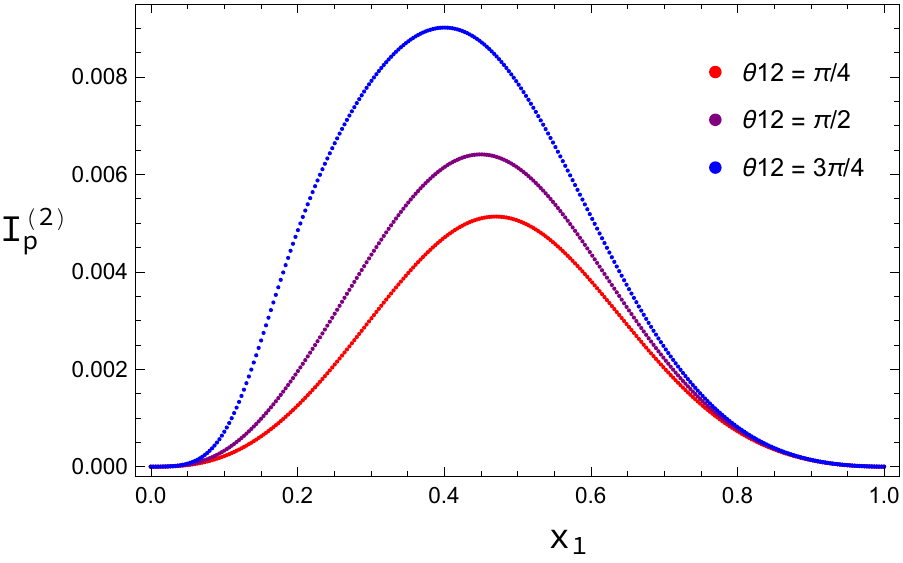}}
\parbox{75mm}{\includegraphics[scale=0.7]{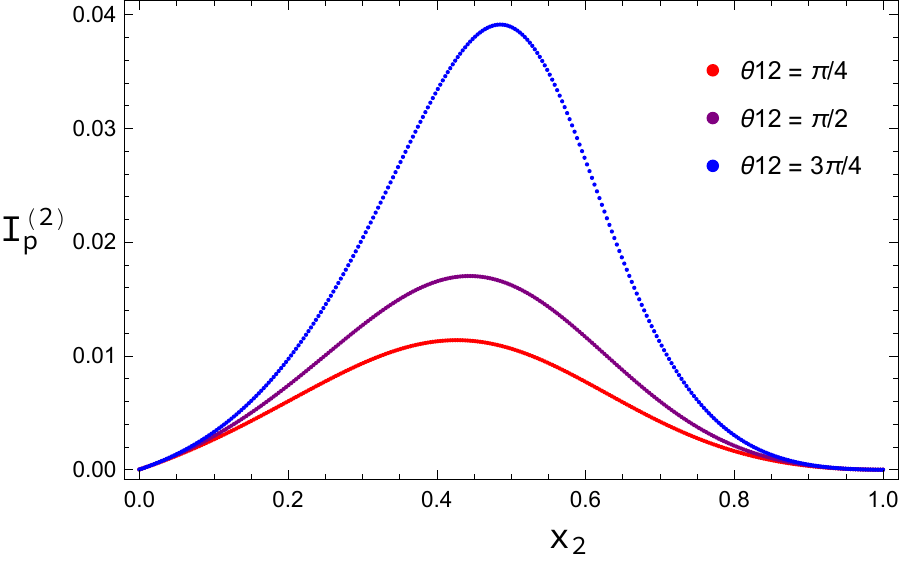}}\vspace{3ex}
(b)\quad\parbox{75mm}{\includegraphics[scale=0.7]{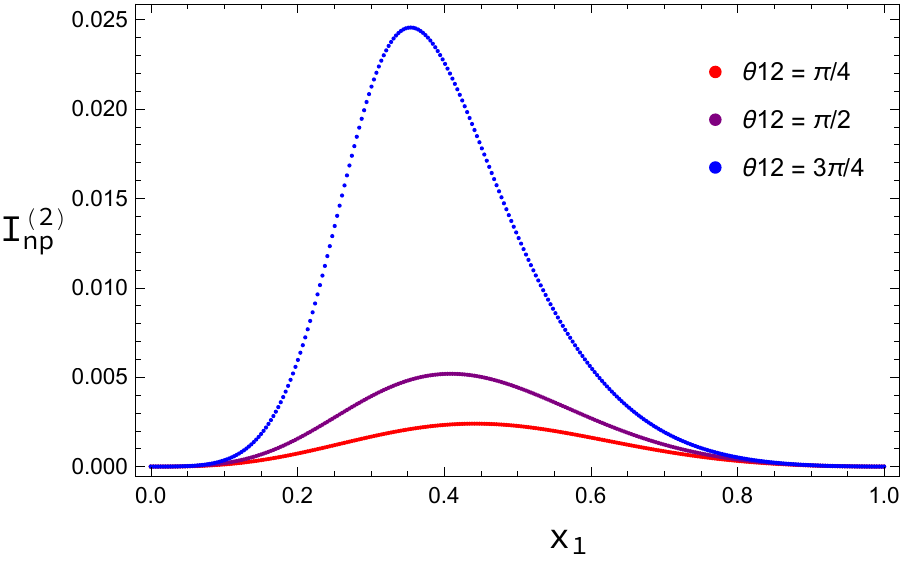}}
\parbox{75mm}{\includegraphics[scale=0.7]{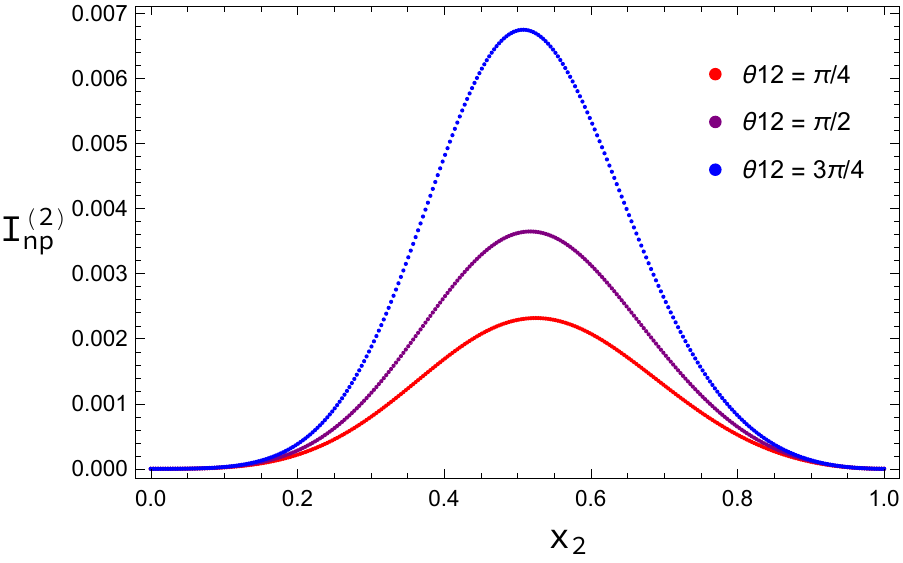}}
\caption{Numerical stability of the integrands of the two-loop (a) planar and (b) non-planar 
scalar triangles of Fig.~\ref{fig:triangles} with random samplings of $x_2$ and $x_1$
in the left and right plots, respectively. 
}
\label{fig:2d-tris}
\end{figure}

In Sec.~\ref{sec:changevars}, we provide the change of variables to spherical 
coordinates leaving the values of radii $r_i$ still unconstrained
and also discuss the mapping to $x_i$, $[0,\infty)\to(0,1]$, 
to avoid the boundary limits. 
With this in mind, we produce, in Fig.~\ref{fig:3d_tris}, a 3D plot 
to show the numerical behaviour
of the planar and non-planar two-loop triangle integrands. 
We vary simultaneously $x_1$ and $x_2$, keeping 
the angles $\theta_{ij}$ fixed. 
Differently to what is carried out in former works~\cite{Aguilera-Verdugo:2020kzc,Ramirez-Uribe:2020hes},
we are not only focused on the structure of the integrand in terms of $\qpw{i}$.
Instead, we provide the complete behaviour of the integrands, where
because of the absence of UV and IR singularities in the latter, 
a numerical integration can straightforwardly performed. 
On top of the 3D plots, we also provide plots in 2D, in Fig.~\ref{fig:2d-tris},
where we scan over $x_1$ ($x_2$) keeping fixed $x_2$ ($x_1$) and varying 
the angle $\theta_{12}$, yet displaying a smooth behaviour at integrand level. 
In other words, the 2D plots in Fig.~\ref{fig:2d-tris} correspond to slices of Fig.~\ref{fig:3d_tris}
for a given $x_i$. 

\begin{table}
\centering
\begin{tabular}{|c|c|c|c|c|}
\hline
\multicolumn{1}{|c}{}& \multicolumn{2}{|c|}{\textbf{Planar triangle}} & \multicolumn{2}{c|}{\textbf{Non-planar triangle}} \\
\hline 
$\frac{s}{m^{2}}$ & LTD ($10^{-6}$) & \SecDec{} ($10^{-6}$) & LTD ($10^{-6}$) & \SecDec{} ($10^{-6}$)\tabularnewline
\hline 
\hline 
$-\frac{1}{4}$ & 9.48(5) & 9.4647(9) & 4.461(3)  & 4.4606(4)\tabularnewline
\hline 
$-1$ & 8.10(5) & 8.0885(8) & 4.101(3)  & 4.1012(4)\tabularnewline
\hline 
$-\frac{9}{4}$ & 6.49(3) & 6.4760(6) & 3.627(5)  & 3.6276(3)\tabularnewline
\hline 
$-4$ & 5.02(2) & 5.0188(5) & 3.15(5)  & 3.1334(3)\tabularnewline
\hline 
$+\frac{1}{4}$ & 10.68(6) & 10.651(1) & 4.743(3)  & 4.7436(4)\tabularnewline
\hline 
$1$ & 13.11(8) & 13.070(1) & 5.259(3)  & 5.2590(5)\tabularnewline
\hline 
$+\frac{9}{4}$ & 20.81(1) & 20.748(2) & 6.533(3)  & 6.5331(6)\tabularnewline
\hline 
$+\frac{25}{16}$ & 15.74(9) & 15.700(1) & 5.748(3)  & 5.7474(6)\tabularnewline
\hline 
\end{tabular}
\caption{Numerical integration with \texttt{NIntegrate} 
of the planar and non-planar two-loop triangles of Fig.~\ref{fig:triangles} 
in the causal representation.
The values of the integrals are compared with \SecDec. 
}
\label{tab:numint}
\end{table}

At this level, we are left with the numerical integration. 
Then, with the \Mathematica{} built-in function 
\texttt{NIntegrate}, we present a coarse integration in which
neither an optimisation nor an educated use of the latter function was
carried out. 
The results of numerical integrations 
of several phase-space points, compared with \SecDec{}, are collected in Table~\ref{tab:numint},
where agreement up-to three digits is found with a simple numerical evaluation. 
The evaluation time per point is $\mathcal{O}\left(30''\right)$ in a 
laptop machine with an Intel i7 (1.2 GHz) processor with 4 cores and 8 GB of RAM.
We provide a \Mathematica{} notebook, \verb"LTD_integrations.wl",
that can be downloaded together with \Lotty{},
with the complete calculation presented in this section. 

We remark that the main purpose of \Lotty{} is not providing a Monte Carlo to numerically
integrate our integrands.
On the contrary, the aim of this \Mathematica{} package is to give a straightforward
understanding of the LTD formalism and, therefore, the causal representation
of multi-loop Feynman integrands. 
In view of the latter, and the various routines to compute the dual integrands 
by means of the Cauchy residue theorem, we claim that any numerator 
written as polynomial in the energy component of the loop momenta 
can be treated along the lines discussed in this paper.
Thus, promoting the causal representation of integrands to 
scattering amplitudes. 
Moreover, it will be desirable to
understand alternative strategies that allow for a more 
stable numerical integration. 

\section{Summary and Outlook}
\label{sec:conclusions}

The novel formulation of the loop-tree duality (LTD)  
allows for an alternative approach to numerically evaluate multi-loop Feynman integrals.
Interestingly, the representation of dual integrands by means of LTD
has been found to display only physical information at integrand level. 
Therefore, the appearance of non-causal contributions, or often called 
pseudo-thresholds, does not generate obstacles when numerically evaluating 
the integrands generated through LTD.

In view of this novel formulation and the results that are being obtained 
in Refs.~\cite{Verdugo:2020kzh,Aguilera-Verdugo:2020kzc,Aguilera-Verdugo:2020nrp,Ramirez-Uribe:2020hes}, 
for topologies up-to four loops
and their generalisation to $L$ loops,
it is desirable to provide a stand-alone package that performs all the calculations
summarised in former works. 
To this end, in this paper we presented the \Mathematica{} package \Lotty{}
that automates the novel formulation of LTD and elaborates
on the causal representation of loop topologies. 
In particular, the close formulae for the all-loop causal representation 
conjectured in Ref.~\cite{Bobadilla:2021rmu} were implemented in \Lotty{},
giving, in this way, a support to the results presented in the latter. 

In order to make a connection between the formalism proposed 
in Ref.~\cite{Verdugo:2020kzh}
and the automation provided by \Lotty{}, we concentrated on particular
cases at two and three loops, showing that the conjectures initially 
proposed in Ref.~\cite{Verdugo:2020kzh} and then proven in Ref.~\cite{Aguilera-Verdugo:2020nrp}
can be worked out regardless of the loop order. 
In particular, we discussed the dual representation of two- and three-loop
scattering amplitudes, showing explicitly that this decomposition,
obtained for vacuum-like diagrams, 
is generalised to topologies with higher multiplicity in the external momenta.
In order to illustrate this behaviour, we considered the two-loop double box 
and the three-loop tennis-court diagrams.

Elaborating on the all-loop causal representation conjectured in Ref.~\cite{Bobadilla:2021rmu},
we presented the routines in \Lotty{} that provide a parametric form of the 
integrands in terms of the physical singularities of the topology. 
To ensure that the latter formula is correct, we numerically compared 
the results obtained from the direct application of LTD against  
the causal formulae, finding completely agreement.   
This numerical evaluation was performed by considering samplings 
of random rational numbers, in order not to lose any precision in the evaluation. 

Furthermore, to study the numerical stability of the integrands in the causal representation,
we considered, as examples, the evaluation of two-loop planar and non-planar triangles,
in which the change of variables to spherical coordinates, implemented in \Lotty{},
allowed us to study the numerical stability of these integrands.
In particular, we presented two- and three-dimensional plots displaying their numerical stability. 

Then, in view of the smooth behaviour of these integrands, 
we numerically integrated out our expressions 
in several phase-space points. This integration was naively performed in 
\Mathematica{} through the function \texttt{NIntegrate},
finding agreement with the results provided by the automated code of \SecDec{}.

\smallskip
The automation presented in this paper allows to implement in a simple way
the LTD formulation and generate multi-loop integrands containing only 
physical information. 
In effect, the dual decomposition of topologies up-to four loops,
including presence of external momenta, have been tested without
showing a limitation. 
Likewise, in the generation of the all-loop causal representation of a topology 
given by its features, cusps and edges, \Lotty{} has been able to provide 
causal representation of loop topologies made of up-to fifteen cusps. 

In view of the implementation of LTD and the structure 
of the causal integrands generated by \Lotty{}, one can focus
only on dealing with physical singularities, infrared (IR), ultraviolet (UV) and thresholds, 
and aim at a four-dimensional finite object at integrand level. 
The cancellation of these singularities has been successfully performed at next-to-leading order (NLO) through the 
four-dimensional unsubtraction scheme~\cite{Hernandez-Pinto:2015ysa,Sborlini:2016gbr,Sborlini:2016hat}
and it is under investigation at next-to-next-to-leading order (NNLO). 
In particular, the calculation of IR-safe objects at NNLO started to appear in the 
numerical evaluation  
of three-point scattering amplitudes at two loops~\cite{Driencourt-Mangin:2019aix,Driencourt-Mangin:2019yhu}. 
More applications, including the treatment of IR singularities, have to be considered 
to provide a purely numerical implementation at NNLO.

We expect that the use of LTD together with its automation, 
given in this paper by \Lotty{}, pushes forward  
the calculations of physical observables
that currently are displayed obstacles when trying to compute them
with the standard methods. 
We emphasis that \Lotty{} is ready to be interfaced with publicly available softwares that
perform the generation and calculation of multi-loop Feynman integrands, e.g. 
Refs.~\cite{Mertig:1990an,Nogueira:1991ex,Hahn:2000kx,Vermaseren:2000nd,Kuipers:2012rf,Shtabovenko:2016sxi}.

\section*{Acknowledgments}

We are indebted to Germ\'an Rodrigo and Roger Hern\'andez-Pinto for uncountless discussions
that led to the automation of the loop-tree duality formalism 
and for all the useful feedback on the implementation of the \Lotty{} package.
We also wish to acknowledge stimulating discussions with Johannes Henn, 
Gudrun Heinrich, Tiziano Peraro, Francesco Tramontano, German Sborlini, 
Felix Driencourt-Mangin, Jonathan Ronca, Jesus Aguilera, Judith Plenter, Renato Prisco, 
Andres Renteria and Selomit Uribe.
This work is supported by the COST Action CA16201 PARTICLEFACE.

\bibliographystyle{JHEP}
\bibliography{refs}

\end{document}